# Language Model Perplexity
# Predicts Scientific Surprise and Transformative Impact

Zhen Zhang[1,2,3] and James Evans[2,3,4*]

[1] School of Information Management, Nanjing University, Nanjing, 210023, China
[2] Department of Sociology, University of Chicago, Chicago, 60637, USA
[3] Knowledge Lab, University of Chicago, Chicago, 60637, USA
[4] Santa Fe Institute, Santa Fe, 87501, USA
[*] Corresponding author. Email: jevans@uchicago.edu

**Scientific breakthroughs typically emerge through the surprising violation of established research ideas[1–4], yet quantifying surprise has remained elusive because it requires a coherent model of all contemporary scientific worldviews[5–7]. Deep neural networks like large language models (LLMs) are arbitrary function approximators[8] tuned to consistently expect the expressions and ideas on which they were trained and those semantically nearby. This suggests that as LLMs improve at generating plausible text, so the perplexity or improbability a text sequence would be generated by them[9] should come to better predict scientific surprise and disruptive importance. Analyzing over 2 million papers across multiple disciplines published immediately following the training of five prominent open LLMs, here we show that higher perplexity scores systematically predict papers that receive more variable peer review ratings[10,11], longer editorial delays, and greater reviewer uncertainty. The most perplexing papers exhibit bimodal outcomes: disproportionately represented among the most celebrated scientific achievements[12,13] and also the most discounted[14]. High-perplexity papers tend to be published in journals with much more variable impact factors and receive fewer short-term citations but in prestigious venues that bet on long-term impact[15,16]. They also generate more interdisciplinary engagement[17] portending long-term influence[18], and are more likely to have been supported by speculative funders like DARPA versus the NIH. Interestingly, we find the opposite pattern for humanities research, where the least surprising work is the most celebrated and cited. Our findings reveal that computational measures of corpus-wide linguistic surprise can forecast the reception and ultimate influence of scientific ideas, offering a scalable approach to recognize and generate potentially transformative research that challenge conventional scientific thinking[19].**

The identification of potentially transformative scientific discoveries holds theoretical significance for the science of innovation[1,2,5,20], but also practical importance for researchers striving for impact, funding agencies and policymakers aiming to foster innovation, and companies and investors pursuing appropriable value. Historical analyses reveal that many groundbreaking contributions—from continental drift theory[21] to the discovery of *helicobacter pylori*'s role in peptic ulcers[22]—initially faced skepticism or outright rejection from the scientific community before eventual recognition. This pattern suggests that revolutionary ideas often violate established expectations about what constitutes scientific knowledge[1,3,23].

Traditional approaches to measuring scientific novelty and impact have relied on citation patterns[24,25], collaboration networks[26,27], and expert assessment[28,29], but these methods are inherently retrospective and require years to manifest meaningful signals. More recently, computational approaches have attempted to quantify novelty through semantic analysis of scientific texts[30–32], unusual combinations of prior knowledge[33,34], or deviations from established research trajectories[35,36]. However, these methods typically focus on narrow aspects of novelty rather than capturing the broader cognitive surprise that revolutionary ideas generate among domain experts.

The emergence of large language models (LLMs) trained on vast corpora presents a unique opportunity to address this challenge[37–39]. These models develop implicit representations of scientific knowledge by learning to predict text sequences, making their prediction failures—measured as perplexity—a potential indicator of content that deviates from established scientific discourse[40,41]. Perplexity quantifies how "surprised" a model is by a given text sequence, with higher values indicating further departure from learned patterns, and lower probability of being generated by the focal model[42]. Crucially, because some LLMs are trained on historical scientific literature, with more literature in the most recent years, their perplexity scores likely reflect the same kind of cognitive surprise that human experts experience when encountering genuinely novel ideas[7,43,44].

Recent work has begun exploring how LLMs assess scientific ideas[45,46], but no study has systematically examined whether model perplexity predicts the human reception of scientific papers or their impact on scholarly discourse and thinking. This gap is particularly important given emerging evidence that transformative discoveries often follow a characteristic pattern: initial uncertainty and mixed reception, followed by either adoption or abandonment depending on empirical validation[47,48]. If perplexity can capture this uncertainty, it might provide an early indicator of ideas with transformative potential. Moreover, LLMs and the deep neural network designs on which they are built have been proven arbitrary function approximators[8] tuned to anticipate the structure of contemporary language. For this reason, as such models improve to generate discourse indistinguishable from humans across contexts, cultures, and scales, model perplexity should come to represent the best possible measure of surprise. Finally because perplexity is measured within a generative model, it can be used to generate entailments to surprising but true claims, enabling the automation of scientific and technological "fast breaks."

We illustrate the computational process whereby LLM perplexity is computed in Fig. 1a with a paper about an unexpected solution to the challenge of understanding how materials transform from liquids into solids when cooled or compressed—the authors' mathematical "distortion-sensitive analysis" approach that detects subtle flaws in existing theories. The perpexity of a paper abstract equals the sum of the log probabilities of each word, conditional on the words that came before it, for all words in the relevant document. We demonstrate that LLM perplexity stably represents the surprise associated with scientific meaning and not simply phrasing by progressively replacing synonyms within scientific papers (abstract) using WordNet (see Materials and Methods M2.2). This shows that expressing the same idea in different ways only marginally changes its model perplexity, converging to less than half of one standard deviation from each original statement's perplexity (Extended Data Fig. 1h).

In order to capture the relationship between LLM perplexity and paper reception, we leverage a natural experimental design afforded by the temporality of model training. We analyzed more than two million research papers published immediately after the training cutoff dates of five prominent open-source language models (OLMo-1B, OLMo-7B, Llama-3-8B, DeepSeek-llm-7B, and GPT-2), ensuring that these papers were new to the models and that their perplexity scores reflected true assessments of their disruptive novelty, uncontaminated with memorized content[49]. Our core hypothesis is that if perplexity captures the same cognitive surprise experienced by human experts when encountering genuinely novel ideas, then paper abstracts with higher perplexity should exhibit observable signatures of uncertainty and mixed reception within the scientific community. We operationalized this by examining multiple independent indicators of scientific surprise and controversy: peer review dynamics (including rating variability and reviewer confidence), editorial processing times, and longer-term impact patterns including citation behavior, interdisciplinary reach, and the sources and style of sponsorship.

**Results**

**Perplexity captures scientific surprise through LLM improbability.** We computed perplexity scores for more than 2 million research papers published shortly after training cutoff dates for five open-source LLMs (see Materials and Methods M1, M2.1), finding substantial agreement across models in how they evaluate surprising research (Spearman's $\rho \geq 0.93$, $p < 2.5 \times 10^{-324}$, $df \geq 1,355,619$ for pairwise comparisons among all tested models; see Supplementary Table S1). All models identified meaningful variation in the predictability of scientific text (Fig. 1a). The perplexity distribution across papers reveals a right-skewed pattern (*skewness* $\geq 1.22$, $p < 2.5 \times 10^{-324}$, $df \geq 8,004$ for all datasets and models tested; see Supplementary Table S2), with most papers exhibiting moderate perplexity scores and a smaller fraction manifesting high perplexity, indicating improbability under the model (Fig. 1b and Extended Data Fig. 1a-c). To validate that perplexity meaningfully captures scientific surprise, we examined papers identified by the scientific community as surprising breakthroughs (Materials and Methods M2.3). Papers featured in prestigious year-end scientific breakthrough lists, including *Nature*'s 10 profiles, *Physics World*'s 2024 Top 10 Breakthroughs, and *Chemical & Engineering News*' Fascinating Findings (Fig. 1b and Extended Data Fig. 1a-c), consistently exhibited significantly higher mean perplexities ($t \geq 3.54$, $p < 0.01$, $df = 32$ in Welch's *t*-test for all models tested; see Supplementary Table S3).

Similarly, papers nominated as "surprising" by an expert panel of 11 scientists across multiple disciplines (Materials and Methods M2.3) showed significantly higher mean perplexity compared to those rated as "unsurprising" (Fig. 1b and Extended Data Fig. 1a-c; $t \geq 3.31$, $p < 0.01$, $df \geq 42$ in Welch's *t*-test for all models tested; see Supplementary Table S4), confirming that model prediction difficulty aligns with human expert assessments of scientific novelty. Analysis of language patterns (Materials and Methods M2.3), graphed in Fig. 1c and Extended Data Fig. 1d-f, and listed in Supplementary Table S5-S8 revealed that high-perplexity papers more frequently employed perplexing terms associated with novel methodologies, unexpected findings, and interdisciplinary concepts. Specifically, perplexing terms were significantly more frequent than expected in the high perplexity group (*standardized residual* $\geq 66.28$ in $\chi^2$ test for all models tested; see Supplementary Table S9), whereas non-perplexing terms were significantly more frequent than expected in the Low perplexity group (*standardized residual* $\geq 35.97$), suggesting that perplexity captures substantive rather than merely stylistic differences in scientific communication.

**High perplexity predicts increased peer review uncertainty and evaluation variability.**
Analysis of 297,101 papers across traditional journals (Materials and Methods M2.3) revealed that acceptance variability—both extremely long (top 1%; $OR = 1.23$, $CI = 1.15\text{-}1.33$, $p = 2.22 \times 10^{-8}$, $df = 297,100$ in logistic regression on logged perplexity) and short reviews (bottom 1%; $OR = 1.92$, $CI = 1.80\text{-}2.06$, $p = 3.41 \times 10^{-78}$, $df = 297,100$; see Supplementary Table S10)—became more frequent as perplexity increased (Fig. 2a). This pattern holds robustly across papers with varying abstract lengths and alternative temporal slices of editorial review times (Extended Data Fig. 2a and Supplementary Table S10).

We further examined peer review processes for 8,005 OpenReview conference papers, which revealed a tight relationship between perplexity and reviewer behavior supporting perplexity as an estimate of scientific surprise (Materials and Methods M1.2 and M2.5). Papers with higher perplexity scores consistently received more variable review ratings, with the intra-paper rating disparity capturing the difference between highest and lowest scores assigned by different reviewers to the same paper, increasing across perplexity deciles (Fig. 2b; $t \geq 2.23$, $p < 0.05$, $df \geq 3,053$ in Welch's *t*-test between top and bottom 20% perplexity groups on 2 out of 4 models except Llama-3-8B and DeepSeek-LLM-7B; see Supplementary Table S11). Critically, high-perplexity papers elicited lower (Fig. 2c; $r \leq -0.02$, $p < 0.05$, $df = 8,003$ in Pearson correlation on logged perplexity across OLMo-1B/7B and DeepSeek-7B, but not Llama-3-8B; see Supplementary Table S12) and more extremely low (Supplementary Fig. S1; bottom 5%; $OR \geq 1.54$, $CI = 1.14\text{-}2.45$, $p < 0.001$ in logistic regression on logged perplexity for all models except Llama-3-8B; see Supplementary Table S13) reviewer confidence scores. These patterns remain robust across papers with varying abstract lengths and alternative definitions of extreme confidence (Extended Data Fig. 2b-d).

Detailed linguistic analysis of peer review comments for the 2,196 Open Review papers with comments confirmed that high-perplexity papers systematically evoke expressions of uncertainty and disagreement among expert evaluators. Reviewers evaluating papers in the top 20% of perplexity employed uncertain language—including terms such as "perhaps," "maybe," "likely," and "might"—at significantly higher frequencies compared to reviews of low-perplexity (bottom 20%) papers (Fig. 2d and Extended Data Fig. 3; $\chi^2 \geq 12.56$, $p < 0.001$, $df = 1$ in $\chi^2$ test for all models tested; see Supplementary Table S14). This linguistic signature of uncertainty was accompanied by measurable increases in evaluation variability across multiple dimensions (Materials and Methods M2.5). The standard deviations of both review ratings (Fig. 2e; $\chi^2 \geq 7.64$, $p < 0.05$, $df = 2$ in White's test on 2 out of 4 models except OLMo-1B and DeepSeek-LLM-7B; detailed results of White's test and OLS regression reported in Supplementary Table S15-S16) and review confidence increases with perplexity deciles (Fig. 2f; $\chi^2 \geq 6.68$, $p < 0.05$, $df = 2$ in White's test for all models tested; results of White's test and OLS regression reported in Supplementary Table S15,S17), suggesting that surprising papers not only receive more uncertain evaluations but also generate greater disagreement among reviewers.

Similar patterns emerge in traditional publishing contexts, where articles with higher perplexity showed increased variability in acceptance processing times (Fig. 2g; $\chi^2 = 23.60$, $p < 0.001$, $df = 2$ in White's test on GPT-2; see Supplementary Table S18) and increased variability in the impact factors of journals that ultimately published them across 15 of 16 natural and social science fields representing 93% of all papers in these fields (Fig. 2h and Extended Data Fig. 4a; results of White's test and OLS regressions for all models and fields reported in Supplementary Tables S19-S20), suggesting that surprising work creates lasting uncertainty regarding its appropriate venue and impact. Interestingly, arts and humanities papers show the opposite trend: JIF variability declines with higher perplexity for all models (Fig. 2i and Extended Data Fig. 4b; results of White's test and OLS regression results all models and fields reported in Supplementary Tables S21-S22).

**Relationships between perplexity and assessments of research quality.** Analysis of 8,005 OpenReview conference papers reveals a positive relationship between perplexity and paper quality. Average ratings increased with perplexity across all models (Fig. 3a; $r \geq 0.03$, $p < 0.01$, $df = 7,878$ in Pearson correlation on logged perplexity for all models tested; see Supplementary Table S23). This relationship was driven by high perplexity papers receiving many more extremely positive ratings (Fig. 3b; top 5%; $OR \geq 1.69$, $CI = 1.20-2.82$, $p < 0.01$ in logistic regressions on logged perplexity for all models tested; see Supplementary Table S24) and fewer extremely negative ratings (Fig. 3b; bottom 5%; $OR \geq 0.56$, $CI = 0.39-0.95$, $p < 0.05$ in logistic regression on logged perplexity for 3 out of 4 models except Llama-3-8B; see Supplementary Table S25), creating a characteristic skewed evaluation pattern. Analysis of honored or award-winning papers revealed that award-winning papers had significantly higher perplexity scores than non-award papers across all models (Fig. 3c and Extended Data Fig. 5c; $U \geq 753{,}536$, $p < 0.05$ in Mann-Whitney U test for all models tested; see Supplementary Tables S26; Materials and Methods M2.6).

When we explore the relationships between the funders in their sponsorship of surprising research, we see a strong relationship (Fig. 3d,e and Extended Data Fig. 5d; Materials and Methods M2.6). Among global sources, those with the strongest relationship include U.S. Department of Defense agencies like the Defence Advanced Research Projects Agency (Fig. 3d; DARPA; $OR \geq 2.21$, $CI = 2.03\text{-}4.05$, all $p \leq 1.48 \times 10^{-71}$ in logistic regression on logged perplexity for all models), the Airforce Office of Scientific Research (AFOSR; $OR \geq 2.58$, $CI = 2.39\text{-}4.37$, all $p \leq 8.64 \times 10^{-140}$) and the Office of Naval Research (ONR; $OR \geq 2.64$, $CI = 2.45\text{-}3.91$, all $p \leq 3.32 \times 10^{-139}$). The U.S. National Science Foundation sponsors published work in proportion to its surprise comparable to Canadian and major European agencies, including the National Sciences and Engineering Research Council of Canada, the European Research Council, and the Deutsche Forschungsgemeinschaft (Fig. 3d,e and Extended Data Fig. 5d; NSF: $OR \geq 1.85$, $CI = 1.82\text{-}2.49$, all $p < 2.5 \times 10^{-324}$; NSERC: $OR \geq 1.55$, $CI = 1.49\text{-}2.19$, all $p \leq 4.40 \times 10^{-99}$; ERC: $OR \geq 1.92$, $CI = 1.86\text{-}3.15$, all $p \leq 4.29 \times 10^{-290}$; DFG: $OR \geq 1.63$, $CI = 1.58\text{-}2.73$, all $p \leq 1.29 \times 10^{-206}$). Asian funders, like the National Science Foundation and the National Natural Science Foundation of China, the Japan Society for the Promotion of Science, and the National Research Foundation of Korea manifest a relatively flat relationship between its sponsored work and surprise (Fig. 3e and Extended Data Fig. 5d; NSFC: $0.88 \leq OR \leq 1.12$, $CI = 0.88\text{-}1.14$, all $p < 0.05$; NSSFC: $1.10 \leq OR \leq 1.77$, $CI = 1.01\text{-}1.92$, $p < 0.05$ for 3 out of 4 models except OLMo-1B; JSPS: $0.70 \leq OR \leq 1.04$, $CI = 0.67\text{-}1.07$, all $p < 0.05$; NRF: $0.82 \leq OR \leq 0.90$, $CI = 0.80\text{-}0.93$, all $p < 0.001$). By contrast, the National Institutes of Health manifests a signficant downward-sloping pattern, favoring more expected developments (Fig. 3e and Extended Data Fig. 5d; NIH: $OR \leq 0.57$, $CI = 0.45\text{-}0.58$, all $p < 2.5 \times 10^{-324)}$; see Supplementary Table S28).

**Top journals seek surprising papers that generate higher interdisciplinary engagement but fewer short-term citations.** Analysis of publication patterns revealed a consistent relationship between perplexity, journal prestige, and citation impact that illuminates the academic ecosystem's response to scientific surprise (Materials and Methods M2.7). High-perplexity papers are disproportionately represented in both the most prestigious journals (top 5% by JIF; $OR \geq 1.72$, $CI = 1.70\text{-}1.99$, all $p < 2.5 \times 10^{-324}$ in logistic regression on logged perplexity for all models tested; see Supplementary Table S29), but also the least prestigious venues (bottom 5%; $OR \geq 1.65$, $CI = 1.62\text{-}2.09$, all $p < 2.5 \times 10^{-324}$ in logistic regression on logged perplexity for all models tested; see Supplementary Table S31) in natural and social sciences, creating a bimodal distribution (Fig. 4a,b and Extended Data Fig 6a-c) that mirrors the polarized evaluation patterns observed in peer review. This pattern remains robust after controlling for abstract length, publication month, and research field (Extended Data Fig. 6,7 and Supplementary Table S33). This suggests that surprising papers either breakthrough to elite venues willing to bet on

transformative potential or become relegated to lower-tier journals when they fail to convince mainstream gatekeepers. As perplexity increases, journal impact factor or JIF rises ($r \geq 0.04$, $p < 2.5 \times 10^{-324}$, $df \geq 1,351,529$ in Pearson correlation on logged perplexity for all models tested; see Supplementary Table S34), whereas short-term citations initially increase and then decline sharply ($\beta \leq -0.47$, $p \leq 1.79 \times 10^{-219}$, $df \geq 1,358,752$ in Quadratic regression on logged perplexity; see Supplementary Table S35) in the natural and social sciences (Fig. 4c and Supplementary Fig. S2). Within high-impact journals (top 5% by JIF), papers with greater perplexity showed a characteristic pattern of receiving fewer immediate citations ($r \leq -0.12$, $p < 0.001$, $df \geq 67,662$ in Pearson correlation on logged perplexity for all models tested; see Supplementary Table S36), indicating that surprising ideas require longer to gain acceptance even when endorsed by elite venues (Fig. 4c, Extended Data Fig. 8a-d, and Supplementary Fig. S2). They must grow new, currently nonexistent markets for attention. The relationship between JIF and citation counts weakened considerably for high-perplexity papers, with decreasing correlation coefficients (Extended Data Fig. 8e-h), indicating that traditional metrics of journal prestige become less predictive of immediate citation impact for the most surprising scientific contributions. These high-perplexity papers demonstrate significantly higher rates of interdisciplinary engagement, however (Materials and Methods M2.8; Fig. 4d and Extended Data Fig. 8i,j). Perplexing papers reference more interdisciplinary papers ($IRR \geq 1.332$, $CI = 1.327\text{-}1.631$, $df \geq 1,344,977$, all $p < 2.5 \times 10^{-324}$ in Negative binomial regression of inter/intradisciplinary references on logged perplexity for all models tested; see Supplementary Table S37) and are cited across more disciplines ($IRR \geq 1.181$, $CI = 1.176\text{-}1.321$, $df \geq 1,045,041$, all $p < 2.5 \times 10^{-324}$ in Negative binomial regression of inter/intradisciplinary references on logged perplexity for all models tested; see Supplementary Table S38). This suggests that transformative work draws from and contributes to knowledge integration across traditional disciplinary boundaries.

Interestingly, arts and humanities show the opposite pattern (Fig. 4e,f and Extended Data Fig 9a-f). With higher perplexity, fewer papers appear in the most prestigious journals (top 5%; $OR \leq 0.32$, $CI = 0.21\text{-}0.34$, all $p < 8.76 \times 10^{-210}$ in logistic regressions on logged perplexity for all models tested; see Supplementary Table S39), and more in the least prestigious venues (bottom 5%; $OR \geq 3.55$, $CI = 3.24\text{-}5.62$, all $p \leq 1.20 \times 10^{-249}$ in logistic regressions on logged perplexity for all models tested; see Supplementary Table S41). Consistent with this pattern, both JIF and citation counts correlate negatively with perplexity (Fig. 4g and Supplementary Fig. S3; JIF: $r \leq -0.23$, $p < 2.5 \times 10^{-324}$, $df \geq 57,482$; citations: $r \leq -0.18$, $p < 2.5 \times 10\text{-}324$, $df \geq 57,725$ in Pearson correlation with logged perplexity for all models tested; see Supplementary Table S44-S45). Moreover, high-perplexity papers in the arts and humanities exhibit lower rates of interdisciplinary engagement (Fig. 4h and Extended Data Fig. 9h; reference: $IRR \leq 0.96$, $CI = 0.75\text{-}0.97$, $df \geq 55,189$, all $p < 0.001$ in Negative binomial regression of inter/intradisciplinary references/citations on logged perplexity for all models tested; citation: $IRR \leq 0.97$, $CI = 0.91\text{-}1.00$, $df = 45,353$, $p < 0.1$ for 2 out of 4 models except Llama-3-8B and DeepSeek-LLM-7B; see Supplementary Table S46-S47). These associations demonstrate that motivations and rewards for surprise are a property of science—natural, artificial, and social—and not humanistic scholarship.

**Discussion**

Our findings demonstrate that LLM perplexity serves as a robust computational indication of scientific surprisingness. As LLMs improve in predictive power, these early findings suggest that they will provide unprecedented insight into how genuinely novel ideas are received and evaluated by the scientific community. By carefully accounting for temporal boundaries—analyzing only papers published after model training cutoffs[49]—we show that computational measures of linguistic surprise can anticipate the complex dynamics that characterize transformative scientific discoveries. This temporal precision is crucial because it ensures that perplexity scores reflect authentic prediction difficulty rather than failed memorization, enabling us to identify genuinely surprising content before its impact becomes evident through traditional bibliometric measures.

**From deductive to abductive AI applications in science.** These results suggest a fundamentally different application of artificial intelligence in scientific discovery than has been pursued to date. Most successful AI applications in science have been essentially "deductive", leveraging existing datasets[50] to interpolate and extrapolate from established facts to unknown instances—exemplified by AlphaFold's prediction of protein structures from amino acid sequences based on patterns learned from experimentally determined structures[51], or drug discovery platforms that identify promising compounds by interpolating within known chemical-biological interaction spaces[52–54]. While powerful, these approaches are inherently conservative, extending established knowledge rather than identifying genuinely surprising departures from it.

Our findings point toward an "abductive" application of AI that could complement deductive applications by identifying and reasoning about unexpected discoveries[7,55]. When a paper receives a high perplexity score, it suggests that the content violates the model's learned expectations about scientific discourse and expectation. Rather than dismissing such content as noise, we can treat high perplexity as a signal of potential paradigm shifts and systematically explore their implications. If we provisionally accept a surprising discovery as true, we can then use reasoning models to explore what other findings would become more probable given this new knowledge, mapping the techno-scientific entailments of paradigm-shifting discoveries like an LLM-based conditional probability. This approach could enable science funders, research institutions, and even AI-driven labs[56] to identify and capitalize on "fast-breaks" that emerge from unexpected discoveries or surprising data, following chains of implication that become visible only after accepting previously improbable premises.

Across the global government funding landscape, our preliminary findings strikingly suggest that U.S. Department of Defense agencies like DARPA have cultivated the consistent capacity to recognize, catalyze and capitalize on paradigm-shifting discoveries that traditional funding bodies overlook or reject. DARPA, AFOSR, and ONR distinctively and disproportionately fund research with high perplexity scores, suggesting that these organizations have developed institutional mechanisms for identifying and supporting transformative work before its impact becomes apparent through conventional metrics[57]. This novel characterization of defense-led funding scientific provides a new justification for establishing DARPA-like agencies for other topics (e.g., ARPA-H for health, IARPA for intelligence, ARPA-E for energy in the U.S.) and in other nations, including the Advanced Research and Invention Agency (ARIA) in the UK, Cyberagentur for cybersecurity in Germany, and Moonshot R&D in Japan. The strategic advantage conferred by early investment in surprising research could prove decisive in technological competition, making the cultivation of institutional risk tolerance and evaluation frameworks that embrace rather than penalize genuine novelty[58] as critical components of national innovation strategies.

**Strategic implications for research investment and evaluation.** The bimodal outcomes we observed for high-perplexity papers—their disproportionate representation among celebrated breakthroughs and ignored obscurities—underscore that surprise alone is insufficient for predicting success. Nevertheless, our temporal approach provides crucial timing advantages that could transform research strategy. By identifying surprising papers immediately upon publication, stakeholders could engage with potentially transformative ideas during their vulnerable early stages, when expert opinion remains divided and future impact uncertain. This early identification capability could enable more strategic resource allocation, allowing funders to place informed bets on high-risk, high-reward research before conventional metrics reveal its potential.

The interdisciplinary citation patterns we observe suggest that surprising discoveries often catalyze knowledge integration across traditional boundaries, creating opportunities for cross-field innovation that may not be apparent to domain specialists. Research institutions could use perplexity monitoring to identify emerging integration opportunities and facilitate collaborations between researchers working on complementary aspects of surprising discoveries. Similarly, the evaluation uncertainty that characterizes high-perplexity papers suggests opportunities for alternative assessment frameworks that embrace rather than penalize genuine novelty.

**Critical methodological considerations and risks.** The success of this approach depends critically on maintaining strict temporal boundaries between model training and evaluation data[59]. Any contamination—where models have been exposed to papers they are supposedly predicting—would fundamentally undermine perplexity as a measure of surprisingness[49]. This data leakage risk will only intensify as language models are continuously updated and as scientific publishing accelerates. Future applications must implement robust versioning and temporal controls to ensure that perplexity scores reflect genuine prediction difficulty.

Additionally, the linguistic focus of our approach may miss surprising discoveries that are primarily methodological, computational, or empirical rather than conceptual. Papers presenting breakthrough experimental techniques or novel datasets might not exhibit high textual perplexity while still representing significant scientific advances. Complementary approaches that analyze methodological novelty, data patterns, or experimental designs could provide a more comprehensive picture of scientific surprisingness.

**Toward anticipatory science policy.** Perhaps most significantly, our findings suggest possibilities for anticipatory rather than reactive science policy. Traditional research evaluation relies on post-hoc measures like citations and awards that reveal impact only years after publication. By providing immediate signals of potential transformation, perplexity-based monitoring could enable more dynamic and responsive research ecosystems. Science agencies could track perplexity patterns across fields to identify emerging areas of high uncertainty and potential breakthrough, adjusting funding priorities to support or redirect surprising developments before their ultimate success or failure becomes apparent.

This anticipatory capability could prove particularly valuable for emerging technologies where societal implications may unfold rapidly. Early identification of surprising technological developments could enable more proactive governance, allowing policymakers to consider regulatory frameworks and ethical implications while innovations remain malleable rather than entrenched. The same temporal advantages that allow strategic research investment could support more thoughtful technology governance, creating space for democratic deliberation about surprising policy ideas and proposals before they become irreversible facts of modern life.

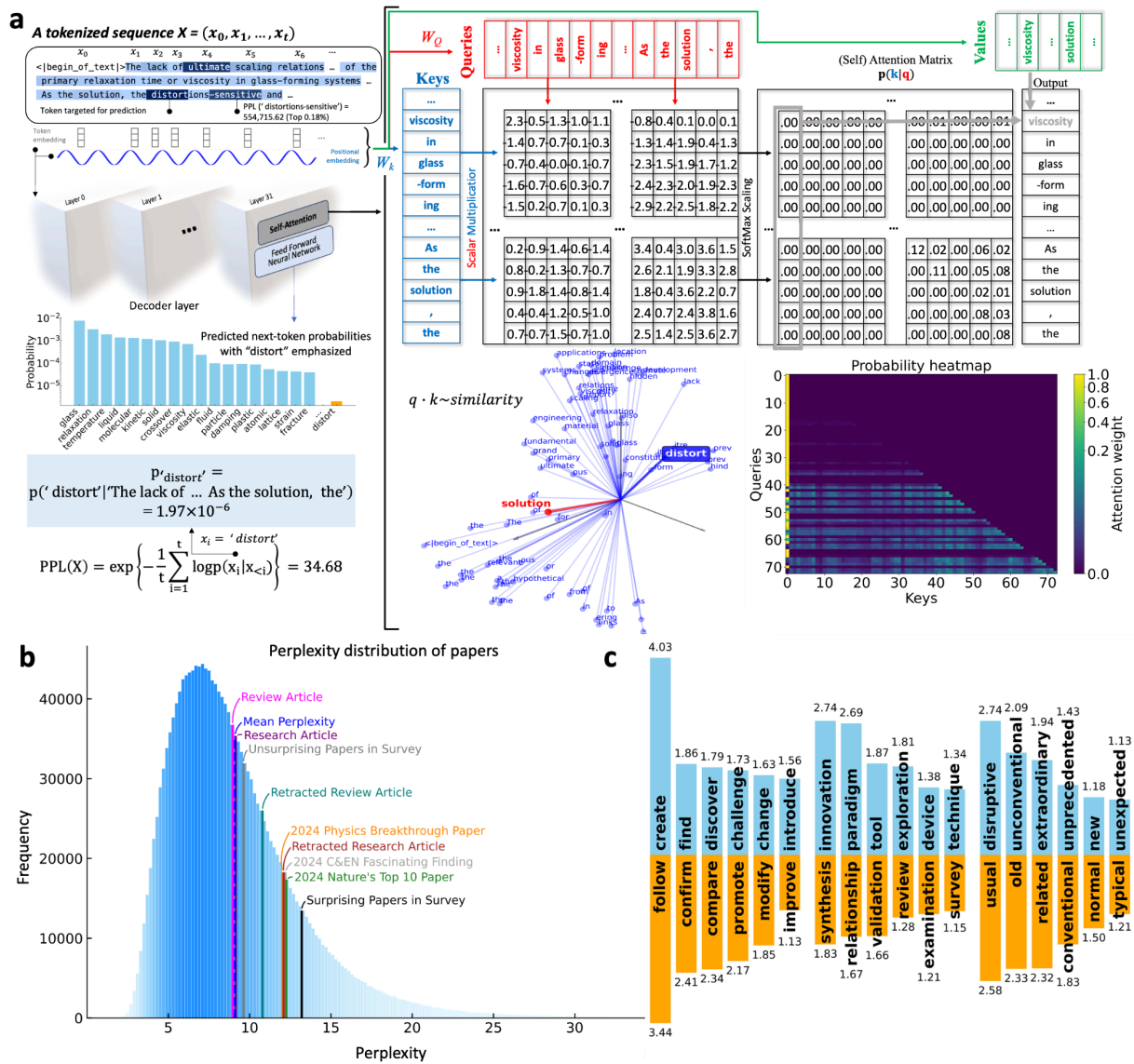

**Fig. 1: Measuring surprisingness via Large Language Models. a**, Illustration of Perplexity (*PPL*) within Large Language Models. Our example input sequence was from the abstract of a material science paper: '*The lack of ultimate scaling relations for previtreous changes of the primary relaxation time or viscosity in glass-forming systems constitutes the grand fundamental challenge, also hindering the development of relevant material engineering applications. The report links the problem to the location of the previtreous domain remote from a hypothetical divergence, hidden in the solid glass state. As the solution, the distortions-sensitive and linearized derivative-based analysis is proposed.*' The model first tokenizes the text and then embeds each token into vector representations before estimating the conditional probability of each token based on its preceding context. In our example, the probability of token 'distort[ion]' (based on Llama-3-8B model) is computed as $p(\text{'distort'}|\text{'The lack of ... As the solution, the'})$. The model's probability of successfully predicting 'distort' is exceedingly low, at merely $1.97 \times 10^{-6}$ and the phrase 'distortions-sensitive' (comprising 'distort', 'ions', and '-sensitive') manifests an average per-token perplexity of 554,715.62, ranking within the top 0.18% of all token-level perplexity values observed in abstracts published in the same journal after the knowledge cutoff date of Llama-3-8B model training. The perplexity of the input sequence is computed as the exponential of the negative

average log probability of each token conditioned on its preceding context. This is derived from the self-attention mechanism, which begins by transforming the token input embeddings into three distinct vector representations: queries ($Q$), keys ($K$), and values ($V$). For each token, its $Q$ and $K$ vectors are multiplied (dot product) such that the following question is answered: "for query token $q$, what preceding keyword tokens $k$ from the sequence best predict it", illustrated by the vector similarities, where "distort" is virtually unrelated to preceding word "solution". This operation produces attention weights, which are transformed via softmax into probabilities that sum to 1, illustrated with the heatmap. These probability estimates are treated as weights and reshaped through multiplication with the token's $V$ vector so that it either generates the next token (if the last attention layer) or forms the basis of the next layer's token embeddings (for every attention layer prior to the last). **b**, Distribution of perplexity, as measured by the Llama-3-8B model on paper abstracts, across 1,784,215 WOS papers in natural and social sciences published after March 2023, the model's training cutoff date. On this distribution, we mark several samples used to validate perplexity. This includes 'surprising' papers (perplexity mean E(*PPL*) = 13.20) and 'unsurprising' papers (E(*PPL*) = 9.14) nominated by a surveyed panel of scholars across fields; (2) 3 groups of papers: one group of papers cited in the Nature's 10: Ten People Who Helped Shape Science in 2024 (E(*PPL*) = 12.25), another group selected among the Top 10 Breakthroughs of the Year in Physics for 2024 by Physics World (E(*PPL*) = 11.02), and a third group highlighted in Chemical & Engineering News' Fascinating Findings of 2024 (E(*PPL*) = 12.19); (3) 1,613,179 original research articles (E(*PPL*) = 9.15), 327 retracted research articles (E(*PPL*) = 12.09), 144,869 review articles (E(*PPL*) = 8.97), and 27 retracted review articles (E(*PPL*) = 10.79). Smaller models exhibit even stronger patterns in the distribution of perplexity—for example, surprising papers tend to occupy more extreme positions—and their results are shown in Extended Data Figs. 1a-c. **c**, After Selecting titles from the 1,784,215 papers and assigning them to one of two groups, perplexing (perplexity in top 50%) or non-perplexing (perplexity in bottom 50%) papers, we calculate the ratio of frequency in perplexing versus non-perplexing papers, $r$. We visualize differences in the type, content, contribution, and value claims between these two groups in terms of verbs, nouns, and adjectives (from left to right). To facilitate comparison, we visualize $r$ in blue if $r > 1$, and $1/r$ in orange otherwise.

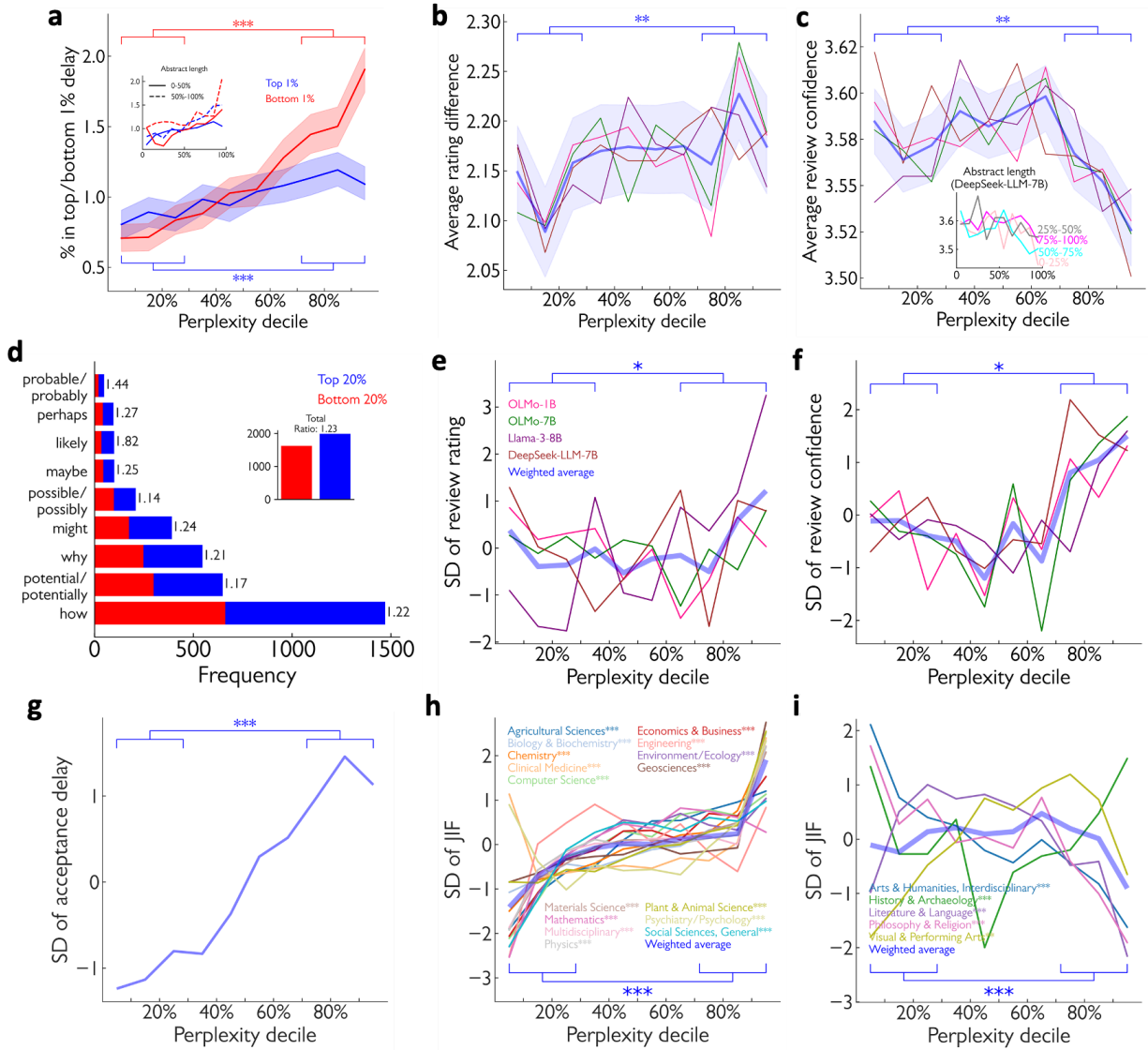

Fig. 2: Validation of perplexity as a measure of surprisingness. **a**, For 297,101 papers published after the end of 2017, the knowledge cutoff date of GPT-2, the proportion of papers with acceptance delays (the difference between received date and accepted date) in the top 1% and bottom 1% both increase with perplexity deciles. Two-sided Chi-squared tests were used between the top three deciles and the bottom three perplexity deciles (*** $p < 0.001$, ** $p < 0.01$, * $p < 0.05$). The inset shows a similar trend across different abstract length groups. **b-c**, For 8,005 OpenReview conference papers visible after May 2023 (beyond the knowledge cutoff of DeepSeek-LLM-7B, the most recently trained model in our study), the average intra-paper rating disparity (difference between the highest and lowest rating scores assigned by different reviewers to the same paper) increases with perplexity deciles, whereas the average review confidence decreases with perplexity deciles across different models (OLMo-1B, OLMo-7B, Llama-3-8B, and DeepSeek-LLM-7B). Welch's $t$-tests were used to compare the means of intra-paper rating disparity and reviewer confidence between the top three and bottom three deciles. The 95% confidence intervals were estimated by bootstrapping all data points within each perplexity decile 1,000 times. The inset in **c** shows the average review confidence decreases with perplexity deciles across groups with varying abstract lengths. For each paper reviewed by multiple reviewers, the mean review confidence across all reviewers were used as the paper level review confidence. **d**, We analyzed review comments from 8,005 OpenReview conference papers,

categorizing them into two groups—surprising papers (top 20% perplexity) and unsurprising papers (bottom 20% perplexity). For uncertain words observed in both groups, we computed their frequency within each group for each model separately, and then averaged these frequencies across models. Reviewers employed uncertain language more frequently when evaluating surprising papers. **e-f**, For the 8,005 conference papers, the standard deviations of review rating and review confidence increase with perplexity deciles across multiple models. **g**, For 297,101 papers published after 2017, the standard deviation of acceptance delays increases with perplexity deciles. **h**, For 1,784,215 Web of Science (WOS) journal papers published after March 2023 (the cutoff date for Llama-3-8B) across 16 natural and social sciences fields, the standard deviation of journal impact factor increases with perplexity deciles. These fields were coded based on the classification of papers into 254 categories and the mapping between these categories and 21 groups provided by WOS. Perplexity scores were calculated using the Llama-3-8B model. **i**, For 74,881 WOS journal papers published after March 2023 across 5 arts and humanities fields, the standard deviation of 2-year journal impact factor decreases with perplexity deciles. Perplexity scores were calculated using the Llama-3-8B model. Similar patterns were observed with perplexity calculated by other models in Extended Data Fig. 4. The Levene's test was used to assess the equality of variances between the top three deciles and bottom three deciles, except for panel **e**, which compared the top four deciles versus the bottom four deciles.

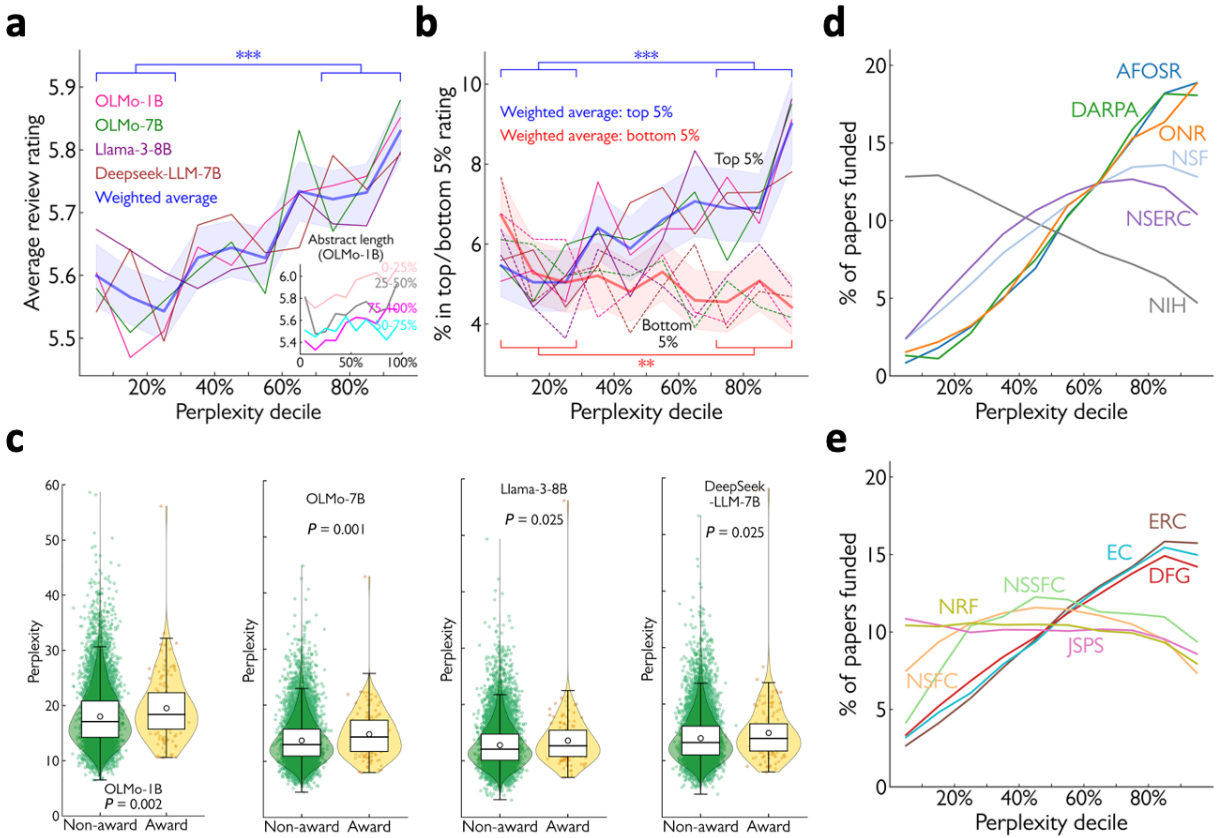

**Fig. 3: The potential relationship between perplexity and quality. a-b**, For 8,005 OpenReview conference papers, the average review rating and the proportion of rating in the top 5% increase with perplexity deciles, whereas the proportion of rating in the bottom 5% decreases with perplexity deciles across different models (OLMo-1B/7B, Llama-3-8B, and DeepSeek-LLM-7B). Welch's *t*-tests were used to compare the means of rating between the top three and bottom three perplexity deciles (*** $p < 0.001$, ** $p < 0.01$, * $p < 0.05$). The 95% confidence intervals were estimated by bootstrapping all data points within each perplexity decile 1,000 times. The inset in **a** shows the average review rating increases with perplexity deciles across groups with varying abstract lengths. **c**, The boxplot shows the mean, median, interquartile range and whiskers extending up to 1.5 times the interquartile range of 10,766 conference papers (127 award-winning papers and 10,639 non-award papers) published after May 2023, with all data points displayed as dots. Award papers demonstrate higher perplexity compared to non-award papers across different models (Mann-Whitney U test, all $p < 0.05$). **d-e**, For 1,784,215 WOS journal papers published after March 2023 (the training cutoff date of OLMo-1B/7B and Llama-3-8B) and 1,358,755 papers published after May 2023 (the training cutoff date of DeepSeek-LLM-7B) across 16 natural and social sciences fields, we identified papers funded by major funding agencies (including NSF, NIH, NSFC, etc.), calculated the proportion of each agency's funded papers falling into each perplexity decile relative to all papers funded by that agency, and then averaged these proportions across 4 models. AFOSR, DARPA, and ONR tend to favor funding surprising research, whereas NIH shows the opposite pattern. Some Asian funding agencies, on the other hand, exhibit a relatively flat relationship with surprise.

## Natural and social sciences

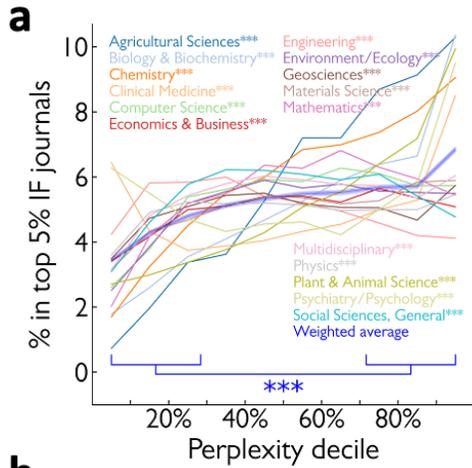

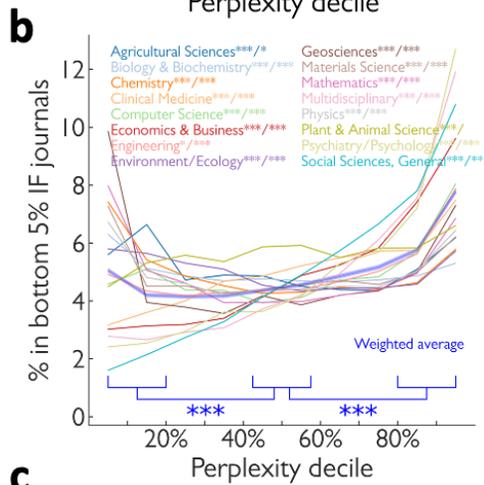

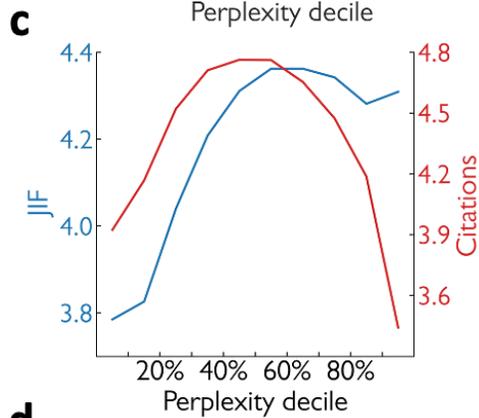

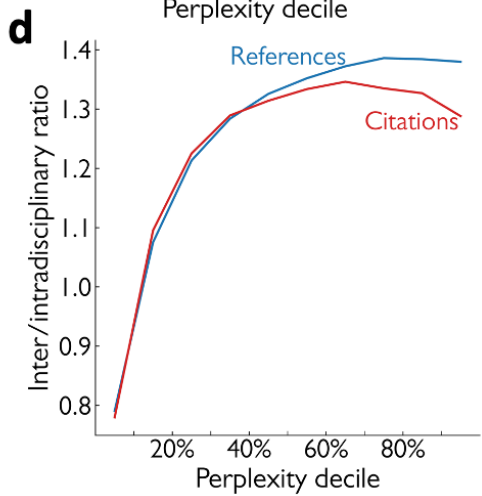

## Arts and humanities

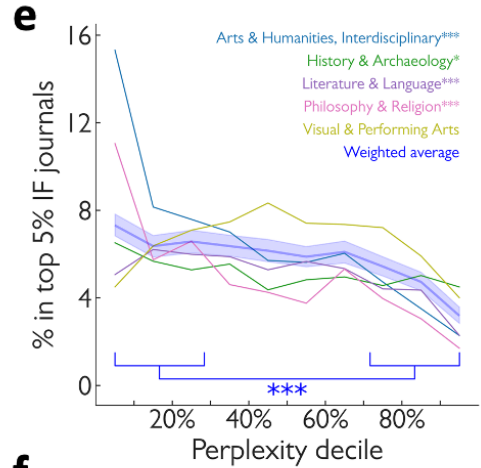

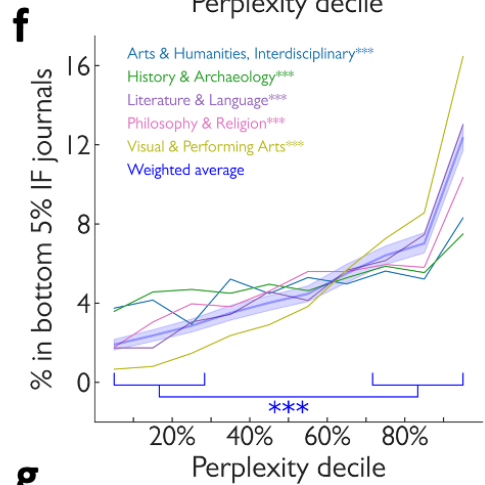

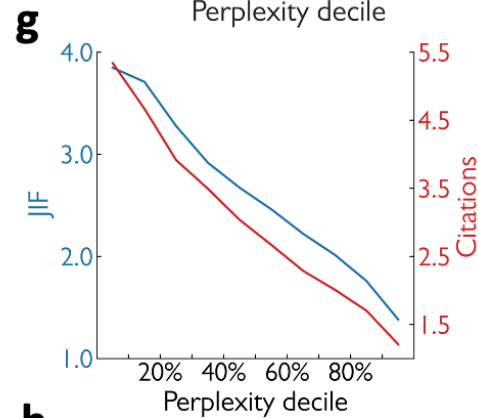

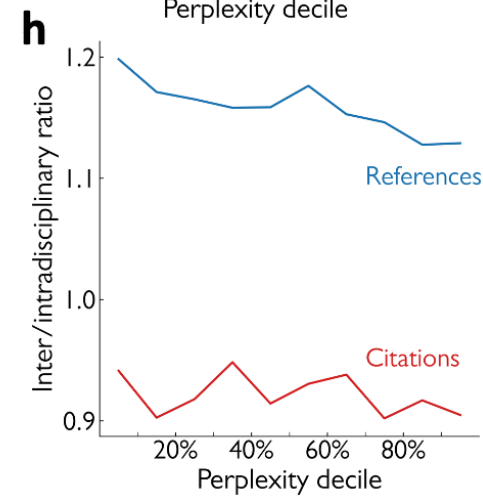

**Fig. 4: Top journals seek surprising papers in the natural and social sciences, whereas in the humanities, surprise is often less valued. a**, For 1,784,215 WOS journal papers published after March 2023 across 16 natural and social sciences fields, the proportion of papers published in journals with a 2-year impact factor in the top 5% increases with perplexity deciles. Perplexity scores were calculated using the Llama-3-8B model. Two-sided Chi-squared tests were used to assess differences in the proportions of papers published in top journals between top 3 and bottom 3 perplexity deciles (*** $p < 0.001$, ** $p < 0.01$, * $p < 0.05$). **b**, As in **a**, but for the proportion of papers published in journals with a 2-year impact factor in the bottom 5%, their proportion increases with perplexity decile. Chi-squared tests were used to assess differences in the proportions of papers published in bottom journals, comparing the top 2 deciles versus the middle 2 deciles, and the middle 2 deciles versus the bottom 2 deciles, given the inverted U-shaped relationship. **c**, The average 2-year impact factor (blue curve, indexed by the left $y$ axis) increases with perplexity deciles, whereas the average citation (red curve, indexed by the right $y$ axis) first increases and then decreases. Both curves are averaged over four LLMs (OLMo-1B/7B, Llama-3-8B and DeepSeek-LLM-7B). **d**, We identified WOS papers' references and citations (78,032,258 references and 7,903,199 citations from 1,784,215 papers for OLMo-1B/7B and Llama-3-8B; 60,370,176 references and 5,642,268 citations from 1,358,755 papers for DeepSeek), which we categorized as either interdisciplinary or intradisciplinary, and computed the ratio of interdisciplinary to intradisciplinary literature. This ratio increases with perplexity for both references and citations. Both curves are averaged over four LLMs. **e-f**, As in **a-b**, but for 74,881 papers across 5 arts and humanities fields. As perplexity decile increases, the proportion of papers published in top 5% impact factor journals decreases, while that in bottom 5% journals increases. **g**, As in **c**, but for 5 arts and humanities fields. The average 2-year impact factor and the average citation both decrease with perplexity deciles. **h**, As in **d**, but for WOS papers from arts and humanities fields. This ratio decreases with perplexity for both references and citations.

## Materials and Methods

### M1 Data Description

**M1.1 Dataset of journal papers.** Our WOS dataset contains 91,322,535 journal papers and 2,357,321,897 citations[60]. To avoid data leakage and accurately assess model perplexity on given paper abstracts, we include papers published after each model's knowledge cutoff date and with valid abstracts. Specifically, for three models (OLMo-1B/7B and Llama-3-8B) with a cutoff date of March 2023[61–63], the dataset contains 1,859,096 papers (a subset of the 2,140,554 published from April 2023 onward with valid abstracts), of which 1,784,215 papers are from 16 natural and social sciences fields and 74,881 papers are from 5 arts and humanities fields. For DeepSeek-LLM-7B, which has a training cutoff date of May 2023[64], the dataset includes 1,416,482 papers (a subset of the 1,637,706 published from June 2023 onward with valid abstracts), of which 1,358,755 papers are from natural and social sciences and 57,727 papers are from arts and humanities. Our analytical conclusions are primarily based on papers from natural and social sciences. For the 1,784,215 papers from natural and social sciences, we identified 78,032,258 references and 7,903,199 citations. For the 1,358,755 papers published after May 2023, we identified 60,370,176 references and 5,642,268 citations, as well as 14,914 journals that published these papers. Subsequently, we identified each natural and social science field paper's references and citations as either interdisciplinary or intradisciplinary. For papers in the arts and humanities for which we computed perplexity using OLMo-1B/7B and Llama-3-8B, we identified 2,126,522 references and 229,302 citations, while papers computed using DeepSeek-LLM-7B yielded 1,681,075 references and 170,164 citations.

The papers that we analyzed are published across 16,570 journals that belong to 254 subfields, according to category labels for journals in the WOS dataset. These 254 subfields can be further classified into 21 fields (i.e., groups in WOS). These 21 fields are (1) Arts and humanities: Arts & Humanities Interdisciplinary, History & Archaeology, Literature & Language, Philosophy & Religion and Visual & Performing Arts; (2) Natural and social sciences: Agricultural Sciences, Biology & Biochemistry, Chemistry, Clinical Medicine, Computer Science, Economics & Business, Engineering, Environment/Ecology, Geosciences, Materials Science, Mathematics, Multidisciplinary, Physics, Plant & Animal Science, Psychiatry/Psychology, and Social Sciences General. Each journal can be assigned to multiple subfields, and each subfield can further be assigned to multiple groups. For each of these journals, we also calculated the 2023 two-year impact factor (Materials and Methods M2.4).

**M1.2 Dataset of conference papers with review information.** The OpenReview dataset contains 27,086 conference papers across 2,065 venues[65]. We collected 8,005 papers that had online publication dates after May 2023, along with the reviewers' rating scores, confidence scores, and comments for these papers. Among them, 2,196 papers include review comments. To compute perplexity scores, we subsequently obtained the abstracts for all 8,005 papers.

**M1.3 Dataset of award and non-award conference papers.** The Semantic Scholar dataset contains 223,962,341 papers published across all fields before 2025[66]. We focus on papers published at seven major AI conferences in 2024: the Conference on Empirical Methods in Natural Language Processing (EMNLP), International Conference on Machine Learning (ICML), European Conference on Computer Vision (ECCV), Computer Vision and Pattern Recognition (CVPR), Annual Meeting of the Association for Computational Linguistics (ACL), AAAI Conference on Artificial Intelligence (AAAI), and International Conference on Learning Representations (ICLR). From these venues, we collected 10,766 conference papers and identified 127 award-winning publications (e.g., best paper awards) among them. To compute perplexity scores for paper abstracts, we subsequently obtained the abstracts for all 10,766 papers.

**M1.4 Dataset of papers with acceptance delay information.** We used a dataset compiled by Liu et al[67]. This dataset contains more than 1,000,000 papers published between 2001 and 2020, along with their submission and acceptance dates. We selected papers with publication dates from 2018 onwards, the training cutoff date of GPT-2, and subsequently obtained their abstracts. In total, our dataset contains 297,101 papers, which we used to examine the relationship between perplexity and acceptance delays.

**M2 Methods**

**M2.1 Calculating perplexity**. Perplexity (PPL) is one of the most common metrics for evaluating language models, which is defined as the exponentiated average negative log-likelihood of a sequence. Since the abstract contains all the key information of a paper, the perplexity we study is calculated based on the abstract of each paper, which serves as a proxy for the paper-level perplexity. Specifically, if we have a tokenized abstract $X$, then the perplexity of $X$ is:

$$PPL(X) = exp\{-\frac{1}{t}\sum_{i=1}^{t} logp_{\theta}(x_i|x_{<i})\}$$

where $logp_{\theta}(x_i|x_{<i})$ is the log-likelihood of the $i$th token conditioned on the preceding tokens[68].

For the WOS dataset (Materials and Methods M1.1), the OpenReview dataset (Materials and Methods M1.2), and the award paper dataset (Materials and Methods M1.3), we computed the perplexity of paper abstracts using the OLMo-1B/7B, Llama-3-8B, and DeepSeek-LLM-7B models. For each model, the papers analyzed were all published after the model's respective knowledge cutoff date. For the acceptance delay dataset (Materials and Methods M1.4), since all papers were published prior to 2020, we only used GPT-2 to compute the perplexity of the abstracts.

**M2.2 Replacing synonyms**. To demonstrate that perplexity can accurately and stably reflect scientific meaning, we adopted a synonym-replacement strategy. Specifically, for papers published after the knowledge cutoff dates of the four models, we randomly sampled 1,000 articles, identified semantically meaningful nouns, verbs, and adjectives, and randomly replaced one such word, recording the change in perplexity. This process was repeated ten times, with the number of replaced words increased by one at each iteration. Synonyms were selected using WordNet, a lexical database of semantic relations between words that links them into semantic networks including synonyms, hyponyms, and meronyms[69]. We find that rephrasing the same idea leads to only minor fluctuations in model perplexity, stabilizing at less than half a standard deviation from the original (Extended Data Fig. 1h).

**M2.3 Validating perplexity.** We conducted four independent analyses to validate perplexity as a measure of surprisingness. Specifically, these include: (1) fielding and analysis of an independent survey of 11 scholars from a range of fields, this survey invited these scholars to propose surprising and unsurprising papers that confirmed our measure; (2) linking papers with breakthrough discoveries recognized by prestigious scientific publications and showing that these nominated works exhibit higher perplexity values, positioning them toward the leading edge of the distribution; (3) extraction and analysis of distinguishing words from paper titles and abstracts that descriptively differentiate perplexing from less perplexing papers; (4) association of paper type and perplexity, revealing that review articles, which synthesizes and summarizes existing knowledge, are less perplexing than original research, and that retracted papers are more perplexing than unretracted papers. We detail each of these in the sections below and highlight each group of papers in Fig. 1b and Extended Data Figs. 1a–c.

*Surprisingness survey*. We fielded an open-ended survey, performed in person or via email. The survey invited scientists across different fields to propose papers they considered either surprising or unsurprising, anchoring the following discussion: 'Surprise in science and technology arises from the violation of expectations held by members of a scientific field about future advances[7]. Our panel of scientists were solicited from 10 prominent research-intensive institutions across the United States, China, United Kingdom, Germany, Australia, Canada, France and South Korea. These scientists came from diverse fields, including atmospheric sciences, bioengineering, computer science, mathematics, medicine, sociology, economics, and management. Among the 11 scientists from whom we received 33 surprising papers and 20 unsurprising papers. The average perplexity of papers nominated to be surprising was higher (OLMo-1B: 21.80, top 9.1%; OLMo-7B: 16.22, top 10.4%; Llama-3-8B: 13.20, top 13.7%; DeepSeek-LLM-7B: 15.59, top 12.4%) than that of papers nominated to be unsurprising (OLMo-1B: 13.12, bottom 55.8%; OLMo-7B: 10.38, bottom 57.4%; Llama-3-8B: 9.63, bottom 64.0%; DeepSeek-LLM-7B: 10.91, bottom 63.2%).

*Breakthrough discoveries perplexity*. We manually collected 33 papers that were mentioned in the following reports: Nature's 10: Ten People Who Helped Shape Science in 2024, Top 10

Breakthroughs of the Year in Physics for 2024, and Chemical & Engineering News' Fascinating Findings of 2024. We then computed the perplexity of these papers. For those mentioned in Nature's 10, the average perplexity ranged from 12.25 to 18.86 (OLMo-1B: 18.86, top 15.8%; OLMo-7b: 14.70, top 15.2%; Llama-3-8b: 12.25, top 17.8%; DeepSeek-LLM-7B: 12.29, top 27.1%). For papers highlighted in the Physics Breakthroughs, the average perplexity ranged from 11.02 to 18.78 (OLMo-1B: 18.78, top 16.0%; OLMo-7b: 13.74, top 19.3%; Llama-3-8b: 11.02, top 25.0%; DeepSeek-LLM-7B: 13.75, top 19.3%). For papers designated as C&EN Fascinating Findings, the average perplexity ranged from 12.19 to 18.33 (OLMo-1B: 18.33, top 17.4%; OLMo-7b: 14.55, top 15.8%; Llama-3-8b: 12.19, top 18.1%; DeepSeek: 14.27, top 17.1%).

*Distinguishing words employed by authors*. We examined the titles and abstracts of papers with high perplexity (top 50%) and low perplexity (bottom 50%) and identified words that appeared frequently in one group but rarely in the other. Specifically, we selected titles and abstracts from our WOS journal papers (1,784,215 for OLMo-1B/7B and Llama-3-8B; 1,358,755 for DeepSeek-LLM-7B) published after March 2023 from natural and social sciences and categorized them into high and low perplexity groups. For words observed in both groups, we calculated the ratio of their frequency in high versus low perplexity papers. We present a sample of popular words with ratios that deviate significantly from 1. These distinguishing words (grouped by part of speech) reflect how papers come to be perceived as surprising. For example, 'create', 'discover', and 'introduce' are among the verbs that most distinguish titles and abstracts of surprising papers, indicating that research introducing new elements tends to be perceived as surprising. In contrast, 'follow', 'compare', and 'confirm' appear more frequently in articles with lower perplexity, indicating expected extensions of existing research. Representative nouns associated with surprising article titles and abstracts include 'paradigm', 'innovation', and 'exploration', which suggests that research challenging existing frameworks or venturing into uncharted territory is more likely to be surprising. Conversely, nouns such as 'relationship', 'review', and 'validation' are overrepresented in less surprising articles, reflecting research focused on consolidating and validating established findings with predictable outcomes. Finally, important adjectives that distinguish surprising article titles and abstracts include 'extraordinary', 'unconventional', and 'unprecedented'. These words suggest that surprising papers frequently investigate underexplored research domains. In contrast, adjectives such as 'old', 'usual', and 'related' characterize work that builds upon extensive prior research. These distinguishing words collectively reveal a clear dichotomy between research that breaks new ground versus research that builds incrementally on established knowledge. The former is more likely to be perceived as surprising, while the latter often yields expected outcomes through incremental advancement.

*Comparison between review articles and original research, and between retracted and unretracted papers*. Review articles focus on existing knowledge, and thus should be less surprising than original research. To validate this hypothesis, we categorize WOS papers into review articles and original research. Among 1,784,215 WOS papers published after March 2023, there are 1,613,179 original research articles and 144,869 review articles (for

DeepSeek-LLM-7B, there are 1,226,310 original research articles and 112,140 review articles among 1,358,755 journal papers). We find that research articles have higher mean perplexity than review articles except DeepSeek-LLM-7B (OLMo-1B: 13.62 vs. 12.27; OLMo-7b: 10.54 vs. 9.96; Llama-3-8b: 9.15 vs. 8.97; DeepSeek-LLM-7B: 11.49 vs. 13.32). An analysis of 354 retracted papers published after March 2023 (for DeepSeek-LLM-7B, there are 186 retracted papers published after May 2023) in the WOS dataset reveals that large language models generally exhibit higher perplexity for retracted papers compared to the mean value of our WOS dataset (OLMo-1B: 15.61 vs. 13.62; OLMo-7B: 12.70 vs. 10.58; Llama-3-8B: 11.99 vs. 9.19; DeepSeek-LLM-7B: 12.06 vs. 10.42). We categorized the retracted papers into retracted research papers and retracted review papers, as shown in Fig. 1b and Extended Data Fig. 1a-c. We also found that as the perplexity decile increases, the proportion of review papers decreases (Extended Data Fig. 5a), while the proportion of retracted papers increases (Extended Data Fig. 5b).

**M2.4 Calculating journals' 2023 impact factor.** To investigate the publication venues of papers with different perplexity levels, we calculated the 2-year JIF for 2023 based on the WOS official methodology[70]. Specifically, for each journal in our analysis, we counted the total citations received in 2023 by all papers published in that journal during 2021 and 2022 (as the numerator) and the number of these papers classified as 'citable items' by WOS (as the denominator). The JIF was then calculated by dividing the total citations by the number of citable items.

**M2.5 Evaluation variability of surprising papers**. Surprise in science and technology arises from the violation of expectations held by members of a scientific field about future advances[7]. Consequently, surprising papers should generate greater uncertainty during the review process. To test this hypothesis, we computed the perplexity of abstracts from 8,005 conference papers using four large language models (OLMo-1B/7B, Llama-3-8B, and DeepSeek-LLM-7B) and 297,101 papers (published between 2018 and 2020) with acceptance delays information using GPT-2, and analyzed the relationship between perplexity and reviewer assessments, including confidences, ratings, rating differences, and acceptance delays. For papers with multiple rating and confidence scores, we used the averages across reviewers as proxies for this paper's rating and confidence. Specifically, we employed binned line plots to reveal potential relationships. In Fig. 2a, papers are ordered by perplexity along the $x$ axis and binned into 10 evenly sized groups. We followed the same binning approach in our main analysis to visualize the results.

We also examined the review comments of 2,196 conference papers in the top 20% and bottom 20% of perplexity scores and identified uncertainty-related words. For uncertainty words observed in both groups, we compared their frequency across four language models. Specifically, for each model, we calculated perplexity and divided the dataset into two groups (high vs. low perplexity). For each group, we computed the frequency of uncertainty words, then averaged these frequencies across the four models. Figure 3a presents the results of this analysis. Results from individual models are provided in Extended Data Fig. 3. We also show analogous analyses comparing papers in the top 50% and bottom 50% of perplexity scores in Extended Data Fig. 4, yielding similar findings. We also examined the journals in which these papers were eventually published. First, we computed abstract perplexity using three models (OLMo-1B/7B and Llama-3-8B) for 1,859,096 WOS journal papers (1,784,215 papers from natural and social sciences; 74,881 papers from arts and humanities) published after March 2023, and using DeepSeek-LLM-7B for 1,416,482 WOS journal papers (1,358,755 papers from natural and social sciences; 57,727 papers from arts and humanities) published after May 2023. We calculated each journal's 2023 two-year impact factor based on citation data, then calculated the standard deviation of impact factors within each perplexity decile. Fig. 2h presents results based on papers from 16 natural and social science fields, while Fig. 2i shows results based on papers from 5 humanities fields, with perplexity calculated by Llama-3-8B. Extended Data Fig. 4a,b show similar field-level results using perplexity values from the other models.

We find that higher perplexity is associated with more extreme long and short review times, larger discrepancies across reviewers, lower reviewer confidence, and more uncertainty words used. As perplexity increases, the variability of review confidence, ratings, and acceptance delays all exhibit substantial increases, indicating that high perplexity is associated with greater evaluation variability across multiple dimensions of the scholarly assessment process. Interestingly, the variability of JIF shows an increasing trend in natural and social science fields, but a declining trend in the arts and humanities.

**M2.6 Analyzing the potential relationship between perplexity and research quality.** We first analyzed the relationship between perplexity and review ratings for 8,005 conference papers from OpenReview. We found that review ratings increased with increasing perplexity. We then examined the differences in perplexity values between award-winning and non-award-winning papers. Specifically, we computed perplexity scores for abstracts from 10,639 non-award-winning conference papers and 127 award-winning conference papers (e.g., best paper awards) using four open-source large language models, and visualized the results using box plots (Fig. 3c). We found that across all models, award-winning papers tended to have higher perplexity values than non-award-winning papers (Supplementary Table S26) and papers with higher perplexity are more likely to receive awards (Extended Data Fig 5c and Supplementary Table S27). We also analyzed the funding agencies of surprising papers. We first identified the funding agencies of WOS papers published after the models' cutoff date based on the acknowledgements of each paper. Specifically, among 1,784,215 journal papers in the natural and social sciences, 260,931 were funded by NSFC, 49,242 by NSF, 43,348 by NIH, 29,754 by EC, 25,204 by NRF, 20,990 by JSPS, 19,195 by DFG, 14,273 by ERC, 11,568 by NSERC, 9,040 by ARC, 2,742 by NSSFC, 2,338 by AFOSR, 2,173 by ONR, and 1,028 by DARPA. Similarly, among 1,358,755 journal papers in the same field published after May 2023, 198,963 were funded by NSFC, 35,419 by NSF, 32,251 by NIH, 22,577 by EC, 19,244 by NRF, 15,215 by JSPS, 14,073 by DFG, 10,374 by ERC, 8,486 by NSERC, 6,769 by ARC, 2,121 by NSSFC, 1,674 by AFOSR, 1,622 by ONR, and 744 by DARPA. Analysis of funding agencies revealed that the proportion of papers supported by certain agencies, notably AFOSR, ONR, and DARPA, increased with rising perplexity. In contrast, the share of NIH-funded papers declined significantly (Fig. 3d).

**M2.7 Analyzing journal and peer preferences for surprising papers.** Surprising papers evoke greater evaluation variability. Given this pattern, which types of journals tend to accept such papers? We therefore examined the relationship between perplexity and JIF as well as citations. Specifically, we analyzed 1,859,096 WOS articles published after March 2023 (the knowledge cutoff date of OLMo-1B/7B and Llama-3-8B) and 1,416,482 WOS articles published after May 2023 (the knowledge cutoff date of DeepSeek-LLM-7B). We first examined the proportion of papers with JIF in the top 5% and bottom 5% across perplexity deciles computed by different models, finding that, for papers in the natural and social sciences, both proportions increased with rising perplexity deciles (Figs. 4a,b, based on Llama-3-8B). Papers with high perplexity tended to cluster at both ends of the journal prestige spectrum, appearing more frequently in both the top 5% and bottom 5% of journals by impact factor. To ensure the robustness of our conclusions, we controlled for abstract length, research fields, and publication time, and repeated the analysis. For papers in the arts and humanities, we found different patterns: as perplexity decile increased, the proportion of papers published in top 5% IF journals decreased, while the proportion published in bottom 5% IF journals increased (Figs. 4e,f, based on Llama-3-8B). Results based on other models are reported separately in Extended Data Fig. 9. We then investigated changes in average JIF and average citations across perplexity deciles. For papers in the natural and social sciences, we found that JIF initially increased then gradually decreased with rising perplexity, while citations increased initially then declined rapidly (Fig. 4c and Supplementary Fig. S2). In contrast, for papers in the arts and humanities, both JIF and citations consistently declined as perplexity deciles increased (Fig. 4g and and Supplementary Fig. S3). Using the natural and social sciences papers, we further analyzed the relationship between JIF and citations across different perplexity levels. We found that regardless of JIF level, high-perplexity papers consistently received fewer citations (Extended Data Figs. 8a-d). Moreover, the correlation between JIF and citations significantly weakened with increasing perplexity (Extended Data Figs. 8e-h).

**M2.8 Analyzing references and citations of surprising papers.** Finally, We examined the references and citations of the 1,784,215 papers in the natural and social sciences, categorizing them as interdisciplinary or intradisciplinary literature based on WOS's classification into 21 groups. For a focal paper A and its reference B, we defined B as intradisciplinary if the set of groups assigned to B was a subset of those assigned to A, otherwise as interdisciplinary. Similarly, for a citation C of the focal paper A, we defined C as intradisciplinary if the set of groups assigned to C was a subset of those assigned to A, otherwise as interdisciplinary. Across the 1,784,215 focal papers, we identified 78,032,258 references and 7,903,199 citations. We then divided focal papers into 10 equal-sized bins by perplexity and calculated, for each bin, the ratio of interdisciplinary to intradisciplinary references and citations. Fig. 4d reports the average across the four models for corresponding deciles, while the results for each individual model are shown in Extended Data Figs. 8i,k. Furthermore, we computed, for each focal paper, the proportion of interdisciplinary references and citations, and then averaged these proportions within each decile (Extended Data Figs. 8j,l). We found that both interdisciplinary reference and citation increased with perplexity, indicating that high-perplexity papers facilitate more frequent interdisciplinary interactions. Further analysis shows that the papers citing perplexing papers are more likely to be published in either top- or bottom-impact journals (Extended Data Fig. 8m,n), while the papers that perplexing papers themselves cite tend to be older, less popular, and published in lower-impact journals (Extended Data Fig. 10). We also conducted the same analysis on papers in the arts and humanities and found different patterns compared to the natural and social science fields: as perplexity increases, interdisciplinary engagement declines (Fig. 4h, Extended Data Fig. 9h,i, and Supplementary Table S46-S47).

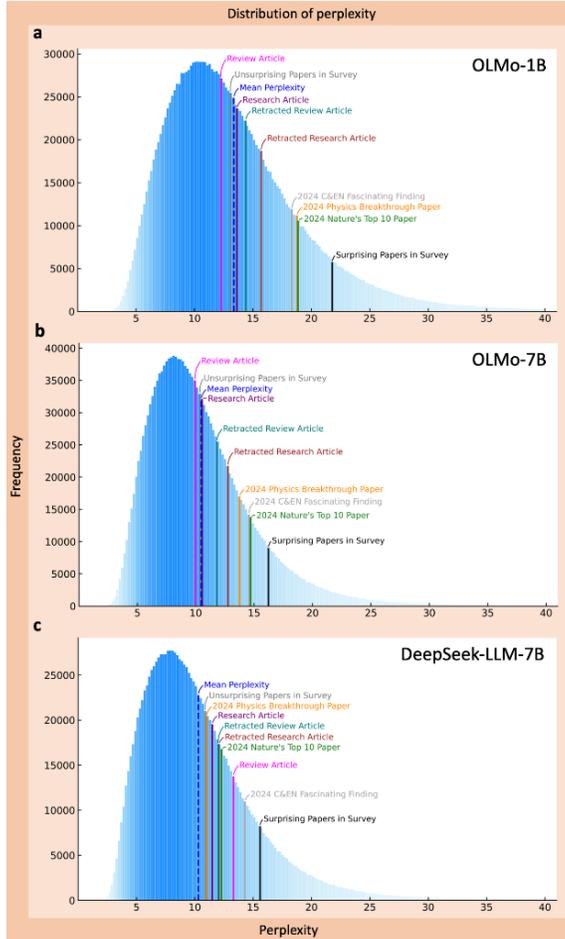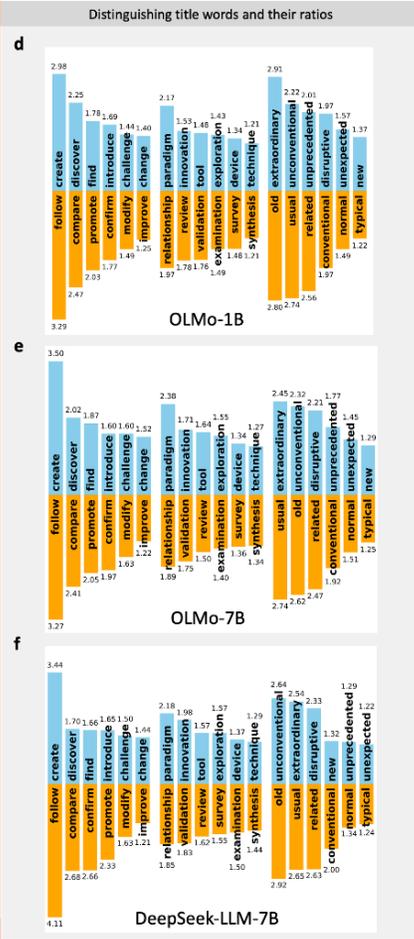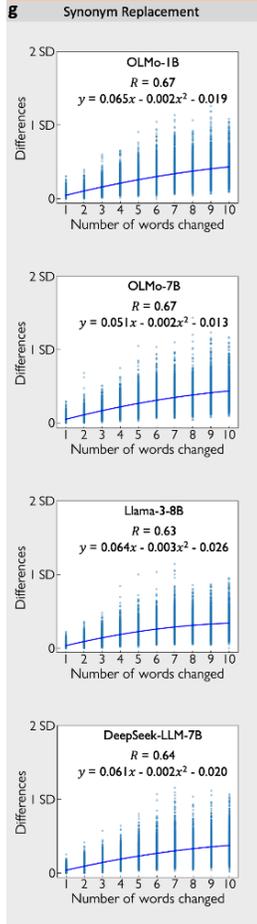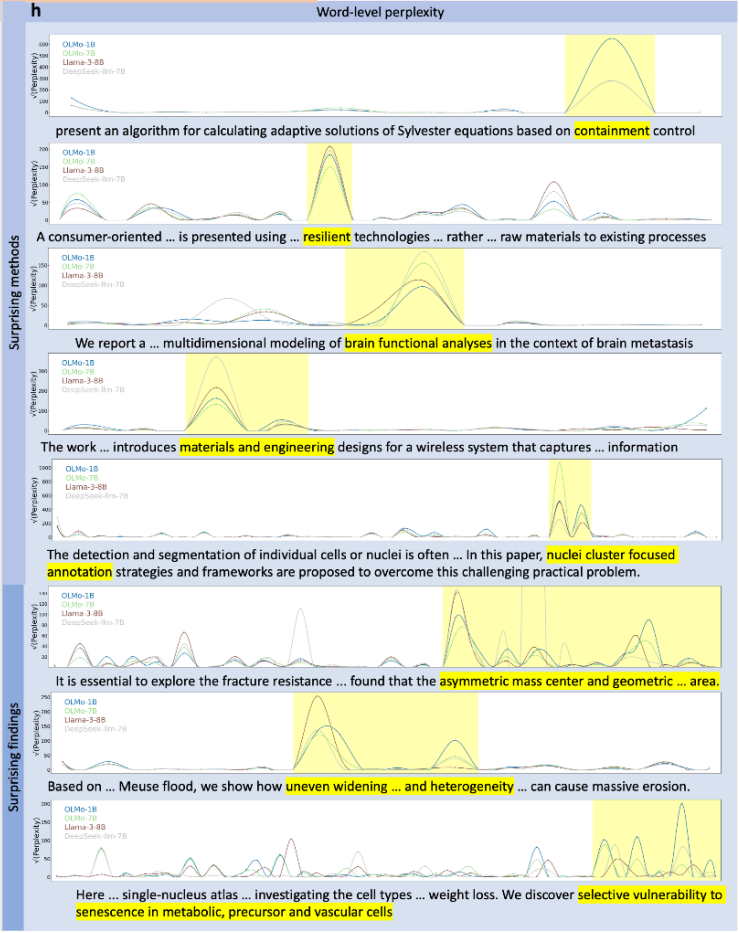

**Extended Data Fig. 1 Perplexity distribution and word-level analysis. a**, We analyzed the distribution of perplexity scores, as measured by the OLMo-1B model on paper abstracts, across 1,784,215 WOS journal papers published after each model's knowledge cutoff date. To validate our perplexity measurements, we examined several paper categories within this distribution. First, expert-nominated papers included 'surprising' publications and 'unsurprising' publications identified by a surveyed panel of scholars across multiple disciplines. We also analyzed papers identified by the scientific community as surprising breakthroughs: Nature's 10 profiles, Physics World's 2024 Top 10 Breakthroughs, and Chemical & Engineering News' Fascinating Findings. Additionally, we examined 1,613,179 original research articles, 327 retracted research articles, 144,869 review articles, and 27 retracted review articles. **b-c**, As in **a**, but perplexity scores were calculated using OLMo-7B and DeepSeek-LLM-7B instead of OLMo-1B. For DeepSeek-LLM-7B, there are 1,226,310 original research articles, 112,140 review articles, 172 retracted research articles, and 14 retracted review articles among 1,358,755 journal papers. **d**, We classified titles from 1,784,215 papers into two groups based on perplexity scores: high-perplexity (top 50%) and low-perplexity (bottom 50%) and examined differences in verbs, nouns, and adjectives to characterize distinctions in type, content, contribution, and value claims. Ratios are displayed in blue when $r > 1$ (enriched in high-perplexity titles) and in orange as $1/r$ when $r < 1$ (enriched in low-perplexity titles). **e-f**, As in **d**, but perplexity scores were computed with OLMo-7B and DeepSeek-LLM-7B rather than OLMo-1B. **g**, To evaluate the stability of perplexity, we randomly selected 1,000 research papers and, in each abstract, randomly identified a semantically meaningful noun, verb, or adjective. We then used WordNet to determine a synonym for the selected word and replaced the original word accordingly. The perplexity of the modified abstract was recalculated. We repeated this procedure 10 times, each time varying the number of replaced words from 1 to 10. The results were visualized using quadratic regression. We found that, consistently across all four models, the absolute change in perplexity increased slowly with the number of replaced synonyms, remained well below the standard deviation of perplexity, and exhibited a negative quadratic coefficient, indicating that perplexity is relatively stable under synonym replacement. **h**, Word-level perplexity trajectories reveal regions of high model uncertainty across 8 paper abstracts, with methods in 5 papers and findings in 3 papers exhibiting higher perplexity values.

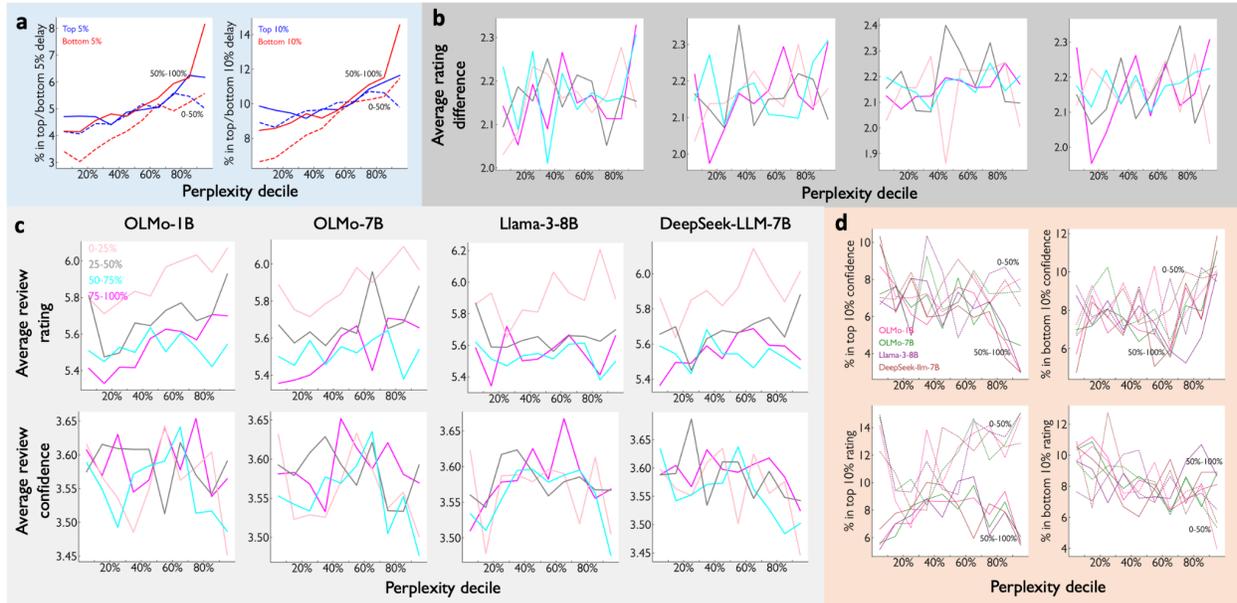

**Extended Data Fig. 2 Examining robustness of perplexity-review information relationship across abstract lengths. a**, We stratified 297,101 papers with available acceptance delay data into two groups based on abstract length (i.e., the number of tokens per abstract, as defined by each LLM's tokenizer) and validated our findings using different thresholds for extreme values (top and bottom 5/10%). We observed that higher perplexity deciles were associated with increased probabilities of both exceptionally long and short acceptance delays. **b-c**, We divided 8,005 conference papers into four equally sized quartiles based on abstract length and examined how intra-paper rating disparity, rating scores, and confidence scores vary with perplexity. Across all abstract length quartiles, increasing perplexity deciles were associated with greater intra-paper rating disparity and higher rating scores, but lower confidence scores. **d**, As in **b-c**, but examining extreme cases rather than average scores. For each abstract length group, we calculated the proportion of confidence scores and rating scores falling into the top and bottom 10% within each perplexity decile. As perplexity decile increased, the proportion of confidence scores in the top 10% and rating scores in the bottom 10% decreased consistently across all abstract length groups, while the proportion of confidence scores in the bottom 10% and rating scores in the top 10% increased.

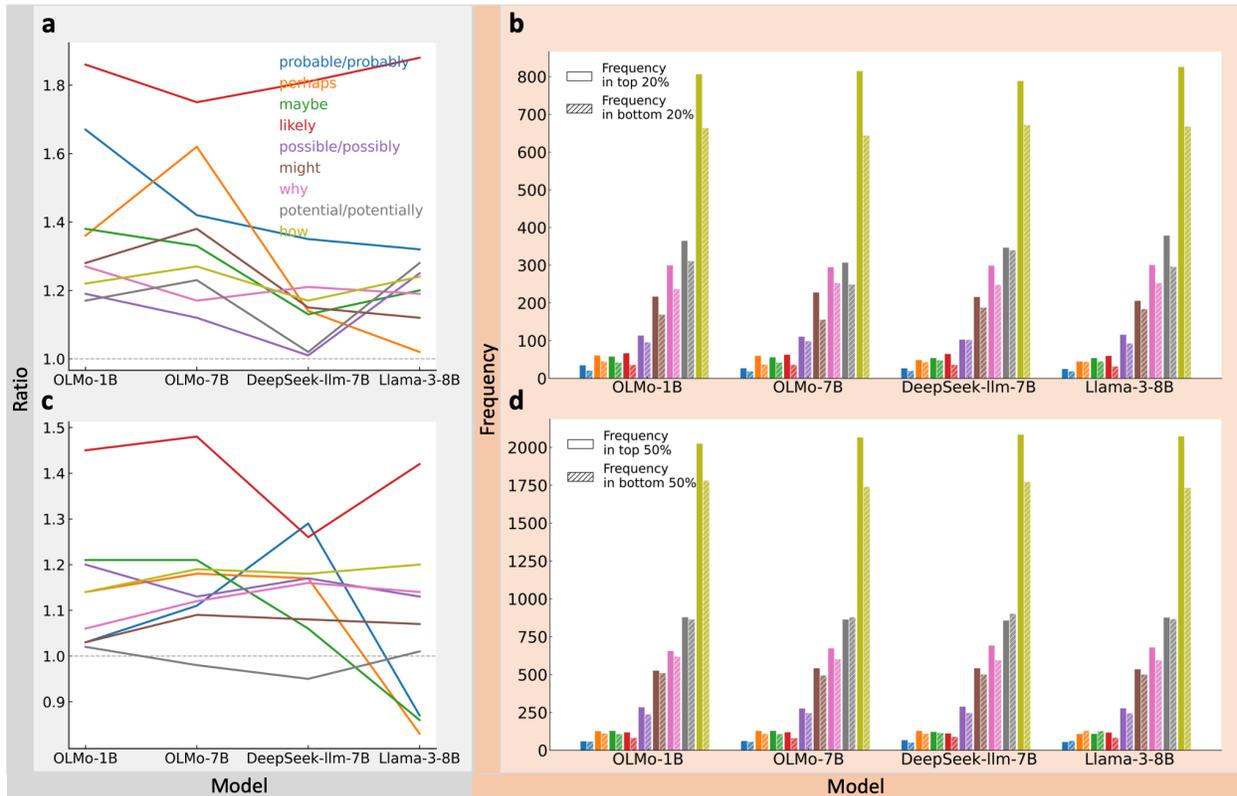

**Extended Data Fig. 3 Uncertain word frequency in review comments grouped by perplexity from different language models. a**, We select 2,196 conference papers with valid review comments and compare the frequency of uncertainty-related words between the highest (top 20%) and lowest (bottom 20%) perplexity groups. We find that words expressing uncertainty (such as perhaps, maybe, and likely), consistently appear more frequently in the high-perplexity group than in the low-perplexity group. Among the models we evaluated, OLMo-1B captures this uncertainty signal most effectively, while Llama-3-8B, despite having the largest number of parameters, performs the worst. **b**, As in **a**, but comparing absolute frequencies rather than relative ratios. **c-d**, As in **a-b**, but comparing the top 50% versus bottom 50% perplexity groups.

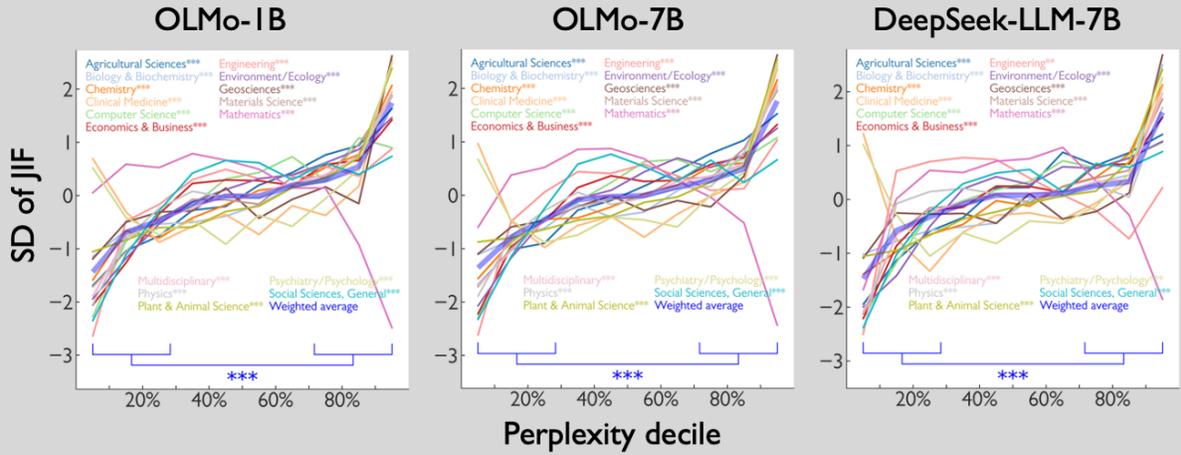
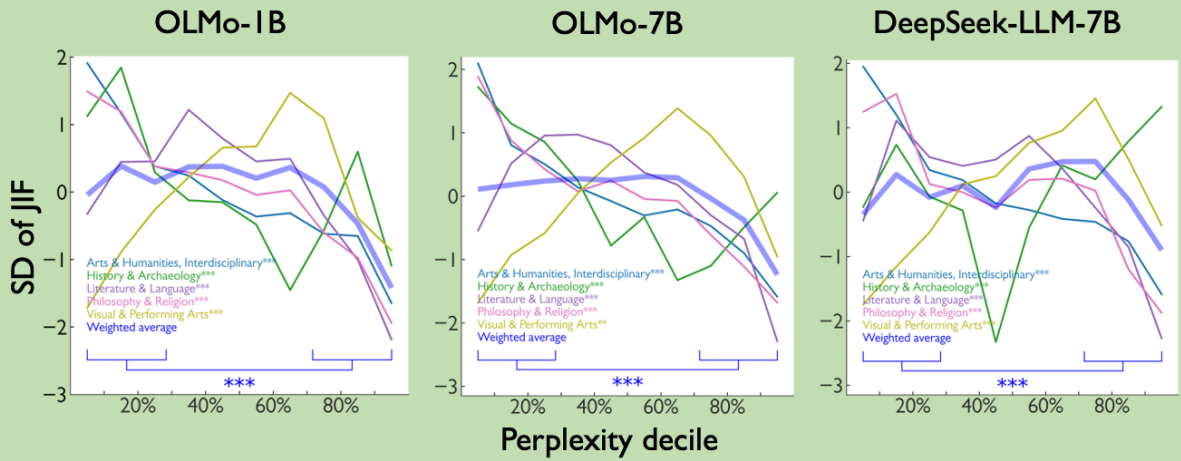
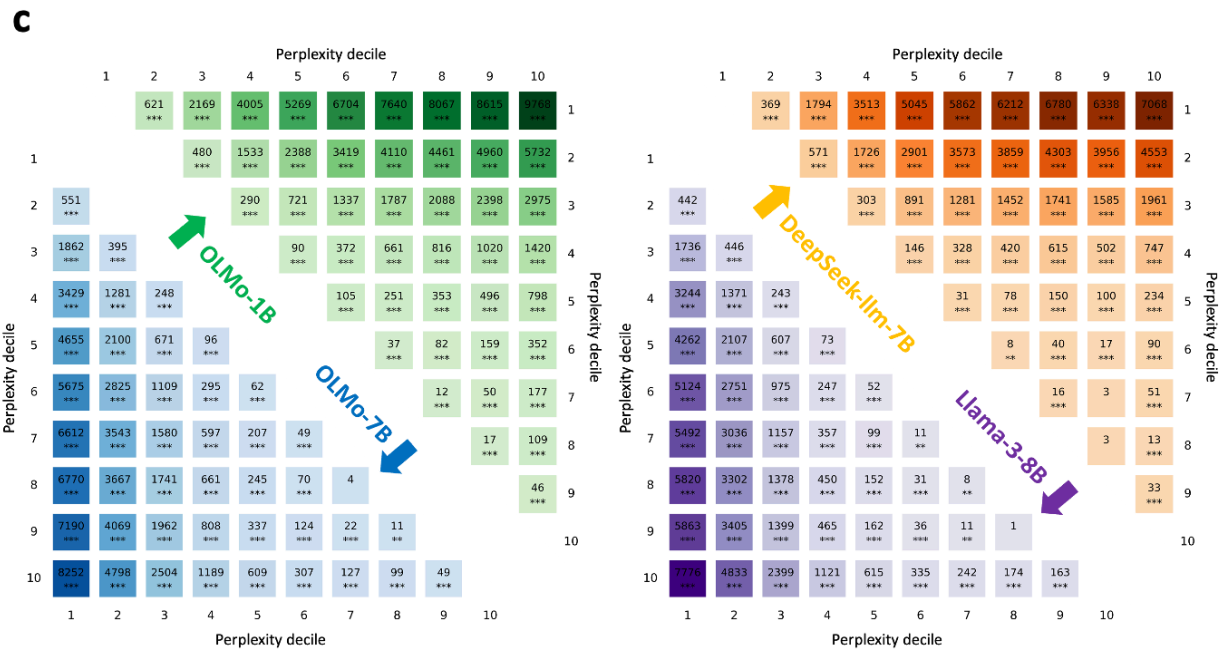

**Extended Data Fig. 4 Assessing variance homogeneity and disciplinary variability of JIF through perplexity decile stratification. a**, In the natural and social sciences, the standard deviation of JIF increases with perplexity decile. For OLMo-1B/7B, we analyzed 1,784,215 journal articles published after March 2023. For DeepSeek-LLM-7B, we used 1,358,755 journal papers published after May of the same year. **b**, As in **a**, but for arts and humanities. In contrast to the pattern observed in the natural and social sciences, the variability of JIF in the arts and humanities tends to decrease with increasing perplexity decile. For OLMo-1B/7B and Llama-3-8B, we analyzed 74,881 journal articles published after March 2023. For DeepSeek-LLM-7B, we used 57,727 journal papers published after May of the same year. **c**, For papers from the natural and social sciences, we tested whether the variance of impact factor differs significantly across different perplexity deciles using the Fligner–Killeen test on all pairwise combinations. The chi-squared statistics are given in the colored cells, and the darkness of the color is proportional to the size of each chi-squared statistic. Asterisks under the numbers indicate $P$ values (*** $p < 0.001$, ** $p < 0.01$, * $p < 0.05$).

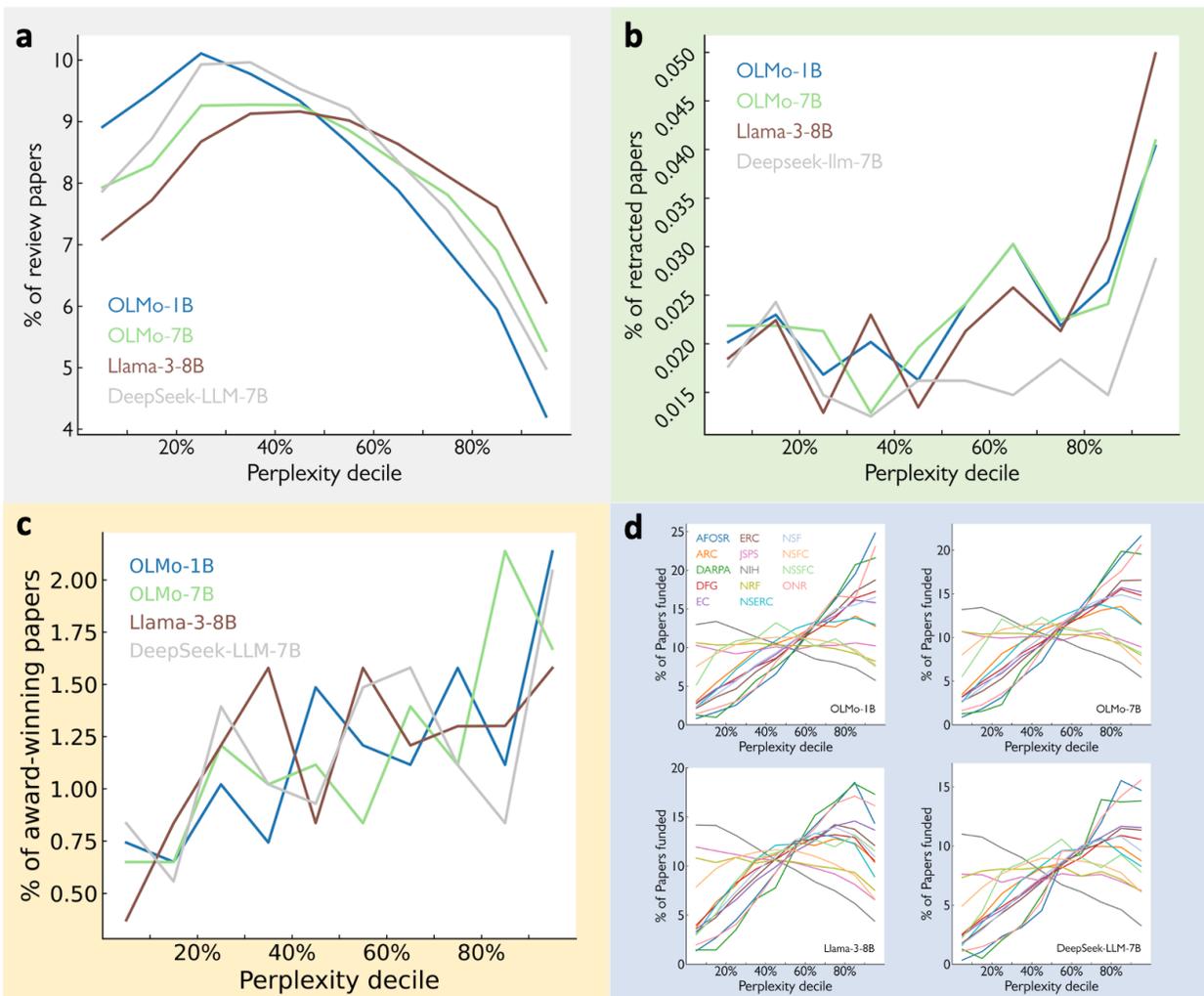

**Extended Data Fig. 5 High-perplexity papers are less likely to be review papers, have higher retraction rates, receive more awards, and are more likely to be funded by agencies such as DARPA that favor high-risk research. a-b**, For 1,784,215 papers published after March 2023, we calculated abstract perplexity using the OLMo-1B, OLMo-7B, and Llama-3-8B models, identifying 144,869 review papers and 354 retracted papers among them. For the DeepSeek-LLM-7B model, we used 1,358,755 journal papers published after May 2023, identifying 112,140 review papers and 186 retracted papers. We found that as the perplexity decile increases, the proportion of review papers decreases while the proportion of retracted papers increases. **c**, For 10,766 conference papers published in 2024 from seven major AI conferences (EMNLP, ICML, ECCV, CVPR, ACL, AAAI, ICLR), we identified 127 papers that received best paper awards or similar honors. We found that as the perplexity decile increases, the proportion of award-winning papers gradually rises. **d**, For the 1,784,215 and 1,358,755 papers mentioned in **a**, we further collected funding acknowledgment information and identified the funder of each paper. We found that as the perplexity decile increases, the proportion of papers funded by certain agencies rises, with the most notable increases observed for AFOSR, DARPA, and ONR. In contrast, the share of papers funded by some other agencies, such as NIH, declines.

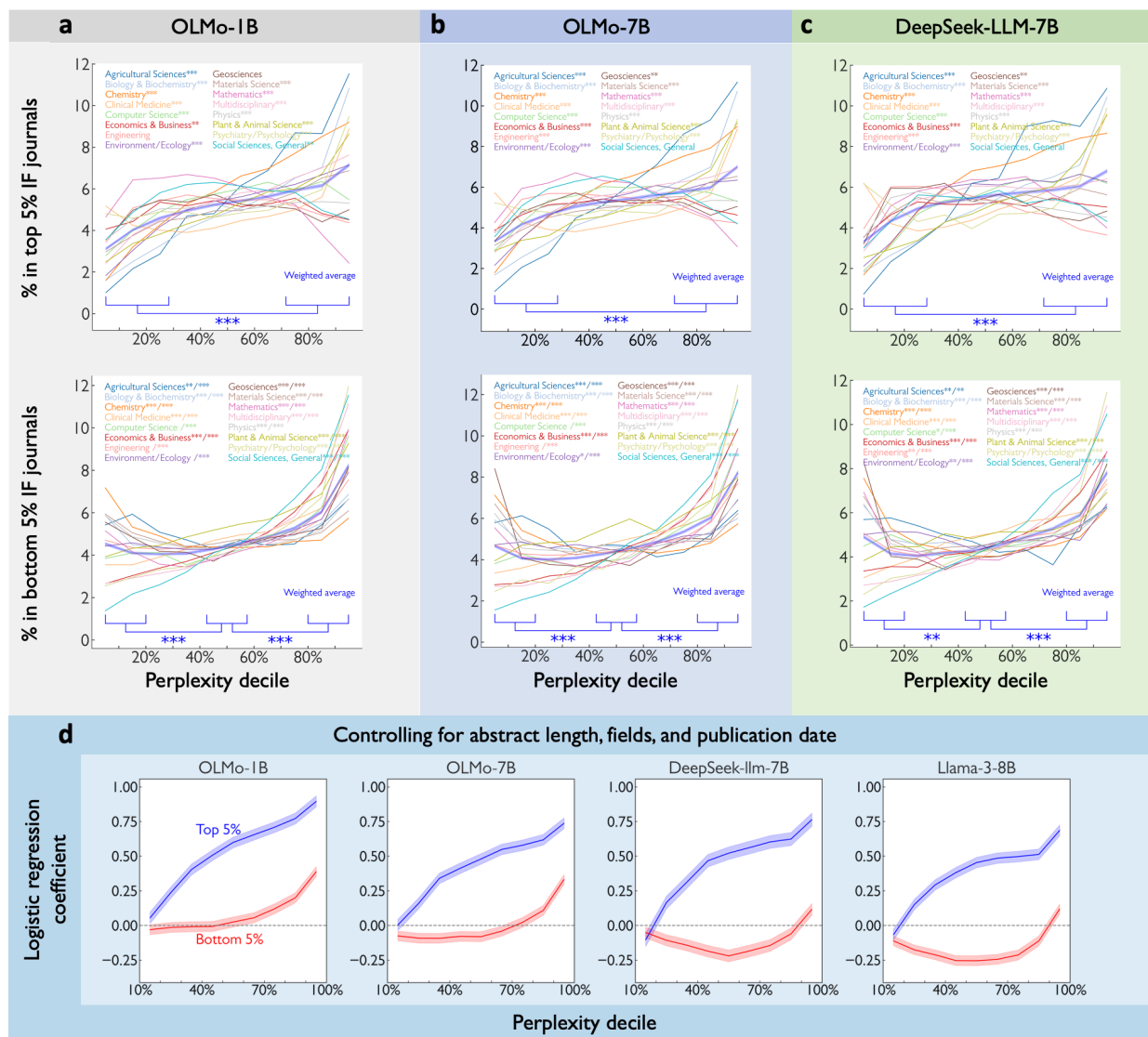

**Extended Data Fig. 6 Increasing top and bottom 5% JIF proportions with perplexity deciles is robust in natural and social sciences across all models. a**, Among the 1,784,215 journal papers published after March 2023 in the natural and social sciences, those that elicit higher perplexity from the OLMo-1B model are more likely to be published in journals ranked in the top or bottom 5% by JIF. **b-c**, As in **a**, but using the OLMo-7B and DeepSeek-LLM-7B models instead of OLMo-1B to compute perplexity from paper abstracts. For OLMo-7B, we used the same dataset described above. For DeepSeek-LLM-7B, the analysis is based on 1,358,755 journal papers from the natural and social sciences published after May 2023. **d**, Plot of the logistic regression coefficients predicting whether a paper was published in a top 5% or bottom 5% journal by JIF, based on perplexity decile, controlling for abstract length, field, and publication date. The regression is based on 1,784,215 journal papers (for OLMo-1B/7B and Llama-3-8B) and 1,358,755 journal papers (for DeepSeek-llm-7B). Field codes from **a–c** were used to control for fields, and abstract length was controlled by token counts derived from each model's tokenizer.

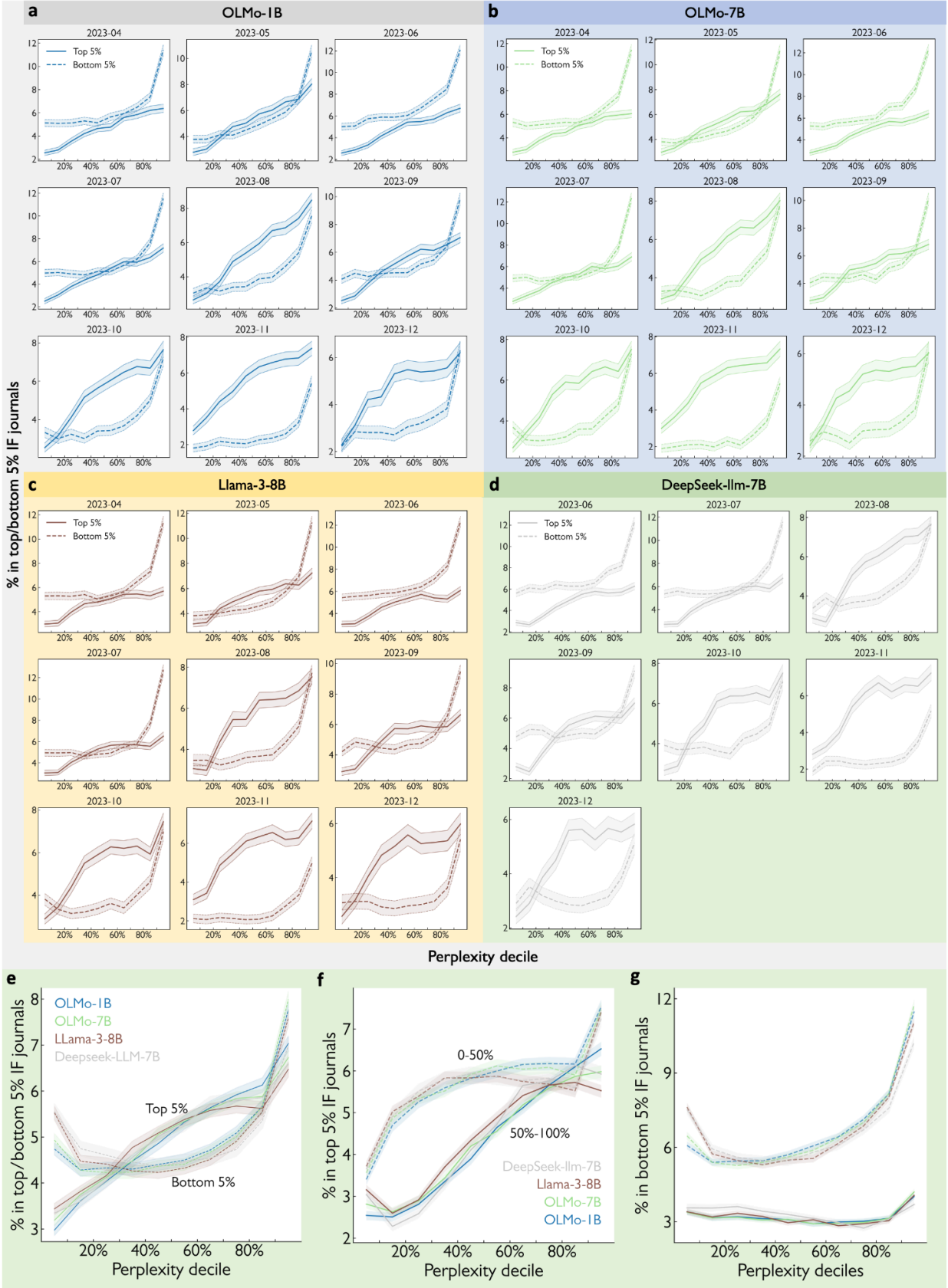

**Extended Data Fig. 7 Perplexing papers exhibit bimodal outcomes when controlling for publication time, abstract length, and using different thresholds. a-d**, For journal papers (OLMo-1B/7B and Llama-3-8B: 1,567,755 WOS papers published between April 2023 and December 2023; DeepSeek-LLM-7B: 1,168,961 WOS papers published between June 2023 and December 2023), the proportions of papers published in top 5% and bottom 5% impact factor journals both increase with perplexity decile. This pattern is robust to variation in publication month within the observed time window and holds across different language models used to compute abstract perplexity. **e**, For WOS papers (1,784,215 for OLMo-1B/7B and Llama-3-8B; 1,358,755 for DeepSeek-LLM-7B), all published after the respective knowledge cutoff dates of the models, ensuring that model outputs were not influenced by prior exposure to the texts, the proportions of papers published in top 10% and bottom 10% impact factor journals are both higher in the highest perplexity decile. **f-g**, As in **e**, but grouping papers by abstract length. Similar patterns were observed: in the highest perplexity decile, papers were more likely to be published in journals ranked in the top 5% or bottom 5% by impact factor.

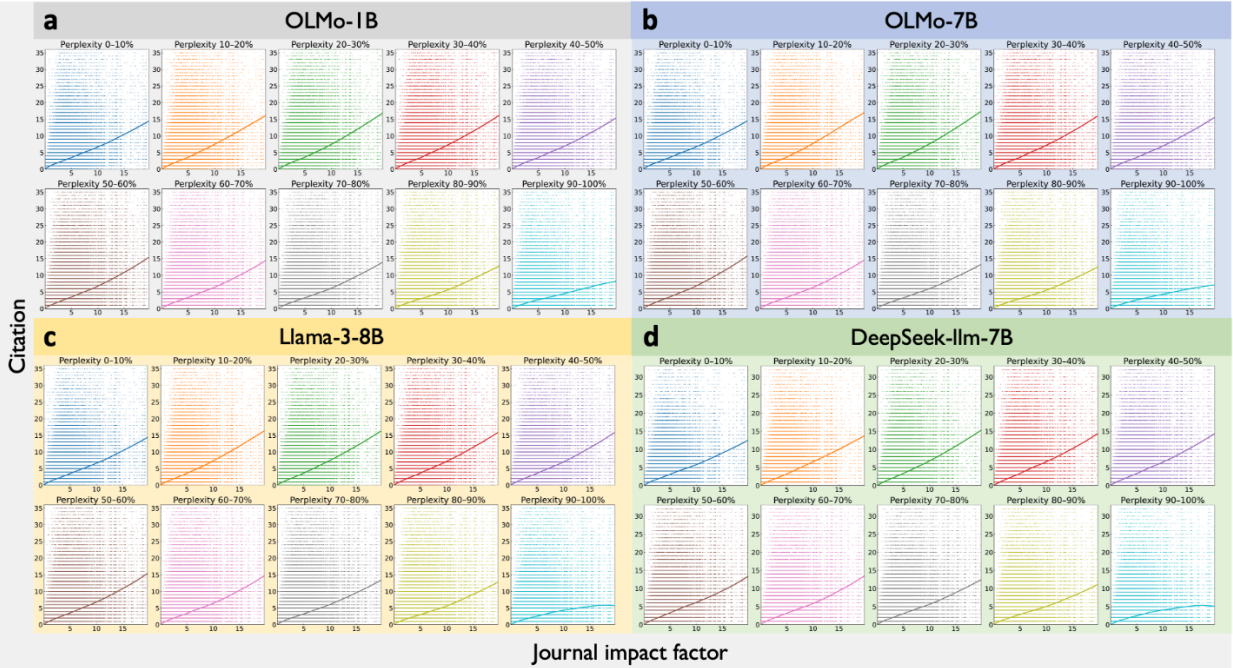
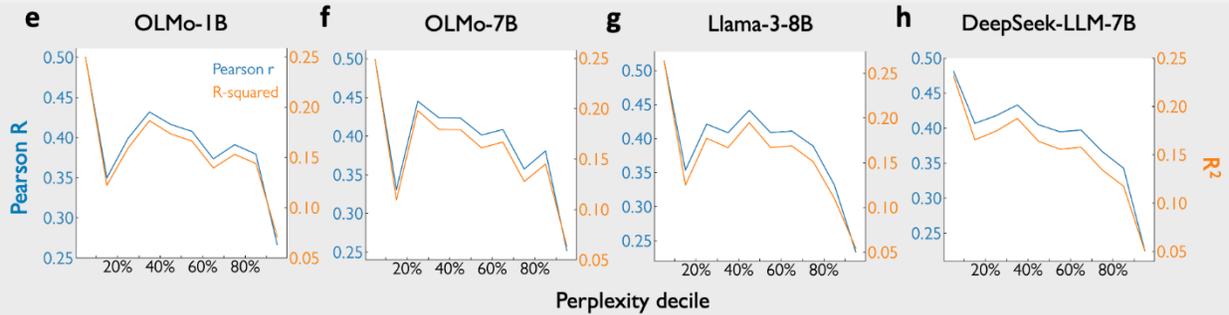
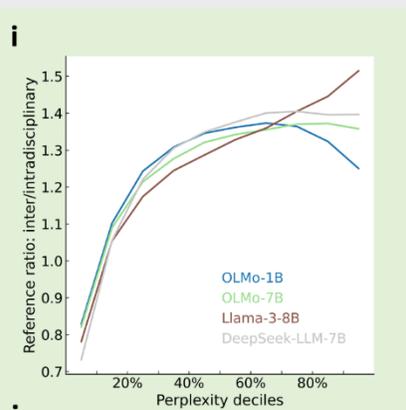
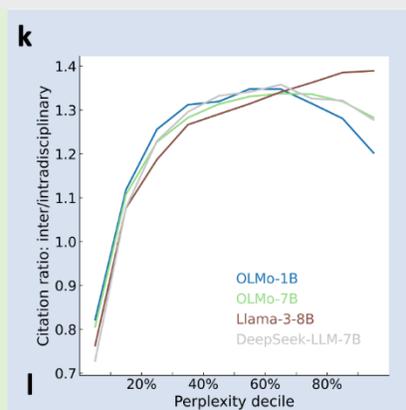
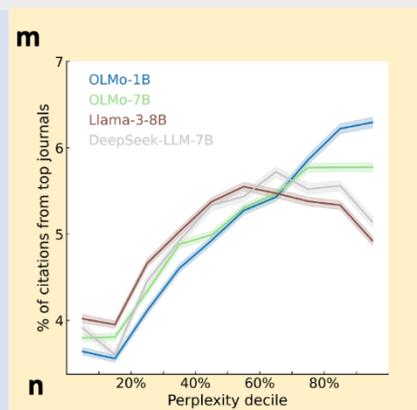
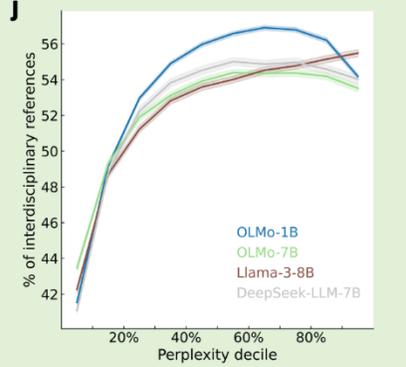
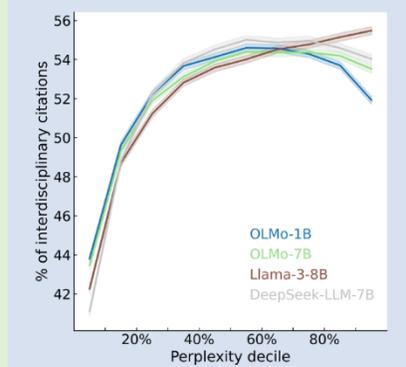
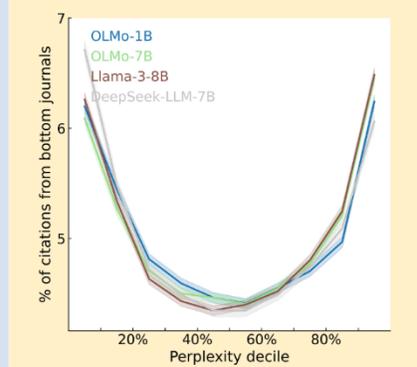

**Extended Data Fig. 8 Perplexing papers consistently receive fewer, yet more polarized, citations, and generate more interdisciplinary engagement. a**, To visualize the relationship between journal impact factor and citation across different levels of perplexity, we first binned articles into ten deciles based on perplexity computed by OLMo-1B. Within each decile, we plotted the number of citations against the JIF for 1,784,215 journal papers published after the knowledge cutoff date of OLMo-1B. We applied LOWESS smoothing, a non-parametric, locally weighted regression method, to capture the underlying trend in each bin. We found that papers within the highest perplexity bin consistently receive lower citation counts compared to those in other bins, and increases in JIF in this bin fail to yield the substantial citation gains observed in others. **b-d**, As in **a**, but using OLMo-7B, Llama-3-8B, and DeepSeek-LLM-7B instead of OLMo-1B to measure perplexity of paper abstracts. **e**, For 1,784,215 WOS papers in the natural and social sciences, we binned these papers into deciles based on perplexity scores (calculated by OLMo-1B) and calculated the Pearson correlation coefficient and coefficient of determination between citation counts and the 2-year impact factor (2023) of their respective journals for each decile. We observed that both the correlation coefficient and coefficient of determination exhibited a declining trend with increasing perplexity deciles. **f-h**, As in **e**, but for OLMo-7B, Llama-3-8B, and DeepSeek-LLM-7B (n = 1,358,755 for DeepSeek-LLM-7B). **i**, We identified 78,032,258 papers cited by 1,784,215 focal WOS papers published after the knowledge cutoff dates of OLMo-1B/7B and Llama-3-8B, and classified these references as either interdisciplinary or intradisciplinary (60,370,176 references cited by 1,358,755 focal WOS papers analyzed for DeepSeek-LLM-7B). Specifically, we used the field labels (groups) described in Materials and Methods M1.1. According to WOS, a single paper may be assigned to one or more groups. For a focal paper A and its reference B, we defined B as intradisciplinary if the set of groups assigned to B was a subset of those assigned to A; otherwise, it was classified as interdisciplinary. We then divided focal papers into ten deciles based on perplexity, ensuring an equal number of focal papers in each decile, though the number of references per decile varied. For each decile, we calculated the ratio of interdisciplinary to intradisciplinary references and plotted this ratio as a line chart. We observed that the ratio of interdisciplinary to intradisciplinary references increases with perplexity deciles. **j**, For each focal paper, we calculated the proportion of interdisciplinary references among all papers it cited. We then averaged these proportions within each perplexity decile and plotted the results as a line chart. We observed that the average proportion of interdisciplinary references increases with perplexity decile. **k-l**, As in **i-j**, but examining the citations received by focal papers rather than the references they cited. Specifically, we identified 7,903,199 citations to 1,784,215 focal papers published after the knowledge cutoff dates of OLMo-1B/7B and Llama-3-8B, and 5,642,268 citations to 1,358,755 focal papers for DeepSeek-LLM-7B. We then classified each citation as either interdisciplinary or intradisciplinary using the same group labels. For a citation C of a focal paper A, we defined C as intradisciplinary if the set of groups assigned to C was a subset of those assigned to A; otherwise, it was classified as interdisciplinary. We observed that papers with higher perplexity tend to be cited from disciplines beyond their own. **m-n**, We further examined the impact factors of the journals that published the papers citing our focal papers and observed that focal papers with higher perplexity were increasingly cited by papers published in journals with top 5% impact factors. Meanwhile, the percentage of citations from journals in the bottom 5% initially declined and then rose. Perplexing papers thus tend to attract more polarized citations, with a disproportionate number of citing articles appearing in both top 5% and bottom 5% journals by impact factor.

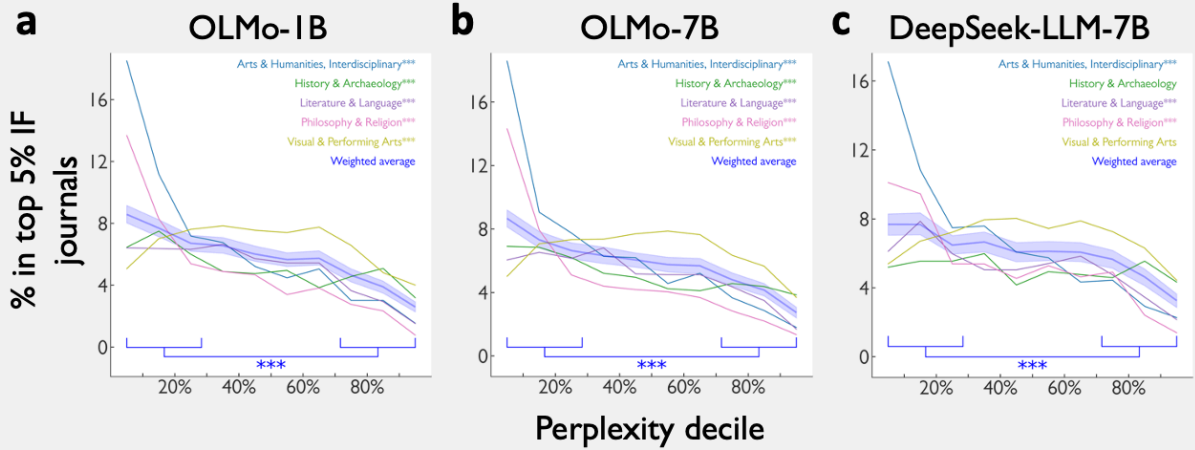
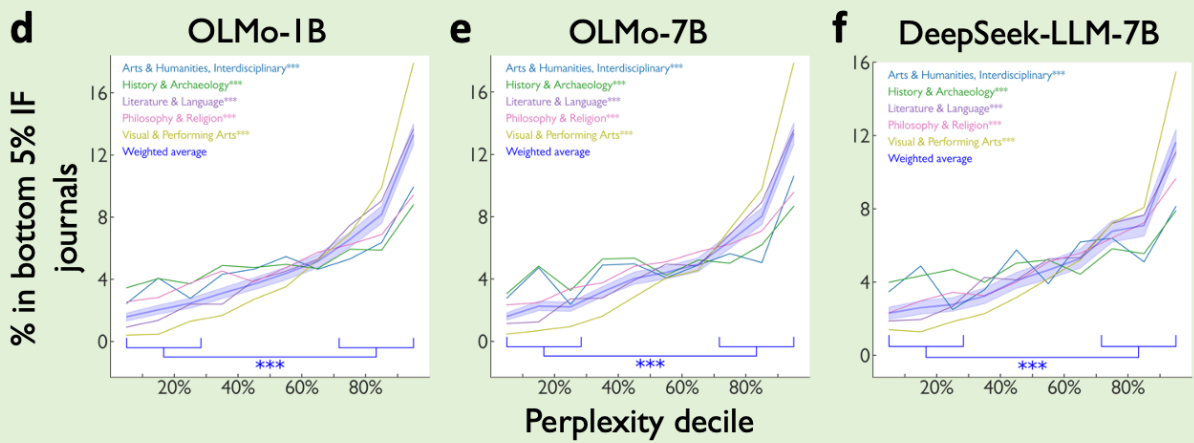
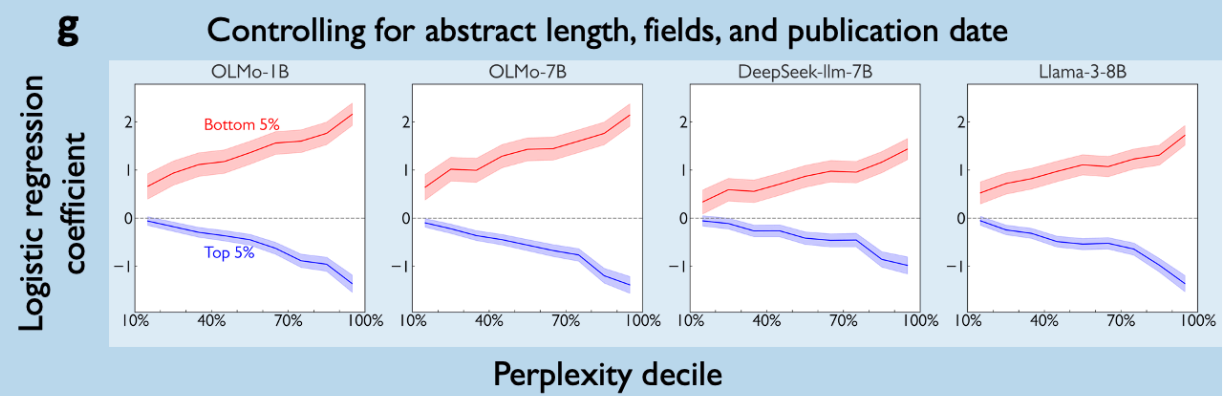
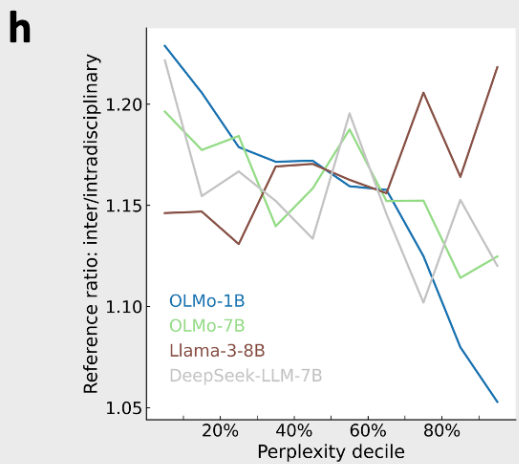
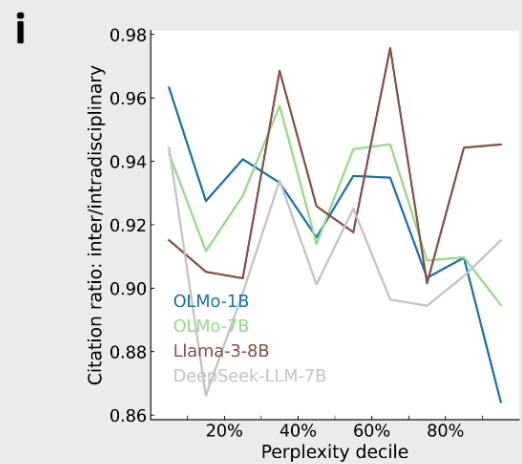

**Extended Data Fig. 9 Perplexing papers are often published in low-impact journals and generate less interdisciplinary engagement in the arts and humanities. a-f**, For journal papers from arts and humanities (74,881 for OLMo-1B/7B, and Llama-3-8B; 57,727 for DeepSeek-LLM-7B), the proportion of papers published in top 5% impact factor journals decreases with perplexity decile, while the proportion in bottom 5% journals increases. This pattern remains robust across different large language models used to compute the perplexity of paper abstracts, as well as across five arts and humanities fields. **g**, Logistic regression coefficients for predicting publication in top 5% versus bottom 5% journals by JIF, as a function of perplexity decile, controlling for abstract length, field, and publication date. Analysis included 74,881 WOS papers (OLMo-1B/7B and Llama-3-8B) and 57,727 WOS papers (DeepSeek-LLM-7B). Field controls used codes from **a-f**; abstract length was controlled using token counts from each model's tokenizer. **h**, We identified 2,126,522 papers cited by 74,881 focal arts and humanities WOS papers published after the knowledge cutoff dates of OLMo-1B/7B and Llama-3-8B, and classified these references as either interdisciplinary or intradisciplinary (1,681,075 references cited by 57,727 focal WOS papers analyzed for DeepSeek-LLM-7B). We divided focal papers into ten deciles based on perplexity and calculated the ratio of interdisciplinary to intradisciplinary references and plotted this ratio as a line chart. We observed that the ratio of interdisciplinary to intradisciplinary references decreases with perplexity deciles. **i**, As in h, but for citations received by focal papers rather than the references they cited, revealing that high-perplexity papers receive fewer interdisciplinary citations.

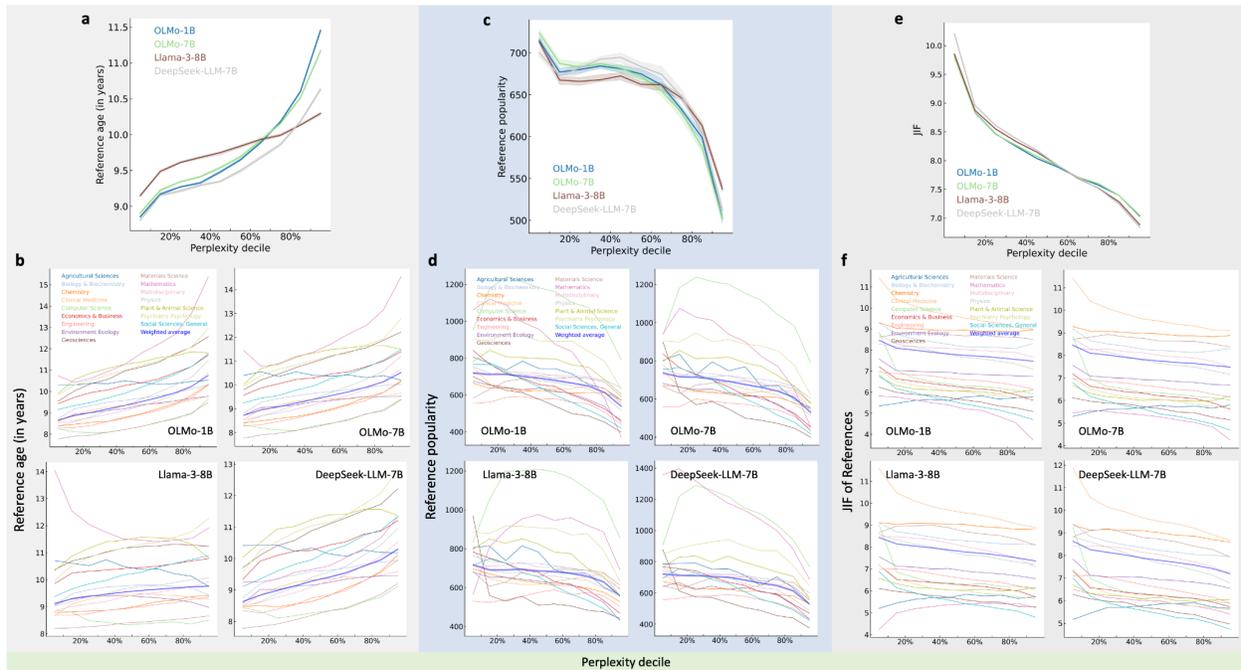

**Extended Data Fig. 10 Perplexing papers build on older, less popular, and less influential ideas. a**, For the references of focal papers identified in Extended Data Fig. 8 (OLMo-1B/7B and Llama-3-8B: 1,784,215 focal papers and 78,032,258 references; DeepSeek-LLM-7B: 1,358,755 focal papers and 60,370,176 references), we computed the reference age as the difference between the publication year of each focal paper and that of its references. We observed that the average reference age increases with perplexity decile of focal papers across four LLMs. **b**, Reference age increases with perplexity decile across 16 natural and social sciences fields, and this pattern holds across perplexity values computed by different LLMs. **c-d**, The average popularity of references (in number of citations) decreases with perplexity deciles across four LLMs and this pattern is robust across 16 natural and social sciences fields. **e-f**, The average impact factor of journals publishing focal papers' references decreases with perplexity decile across four LLMs, and this pattern is robust across 16 natural and social sciences fields.

# Supplementary Information for Language Model Perplexity Predicts Scientific Surprise and Transformative Impact

Zhen Zhang[1,2,3] and James Evans[2,3,4*]

[1] School of Information Management, Nanjing University, Nanjing, 210023, China
[2] Department of Sociology, University of Chicago, Chicago, 60637, USA
[3] Knowledge Lab, University of Chicago, Chicago, 60637, USA
[4] Santa Fe Institute, Santa Fe, 87501, USA
[*] Corresponding author. Email: jevans@uchicago.edu

**The PDF file includes:**

Figs. S1 to S3
Tables S1 to S47

**Figure S1. Percentage of reviews in top 5% and bottom 5% confidence by perplexity deciles.**

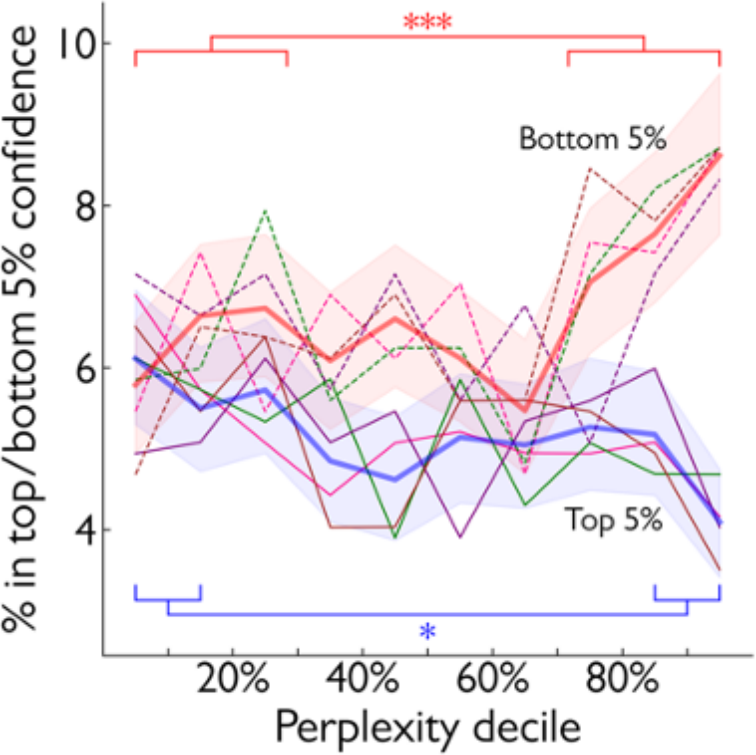

**Figure S2. Trends of average JIF and average citations across perplexity deciles calculated by four different models for papers in natural and social sciences.**

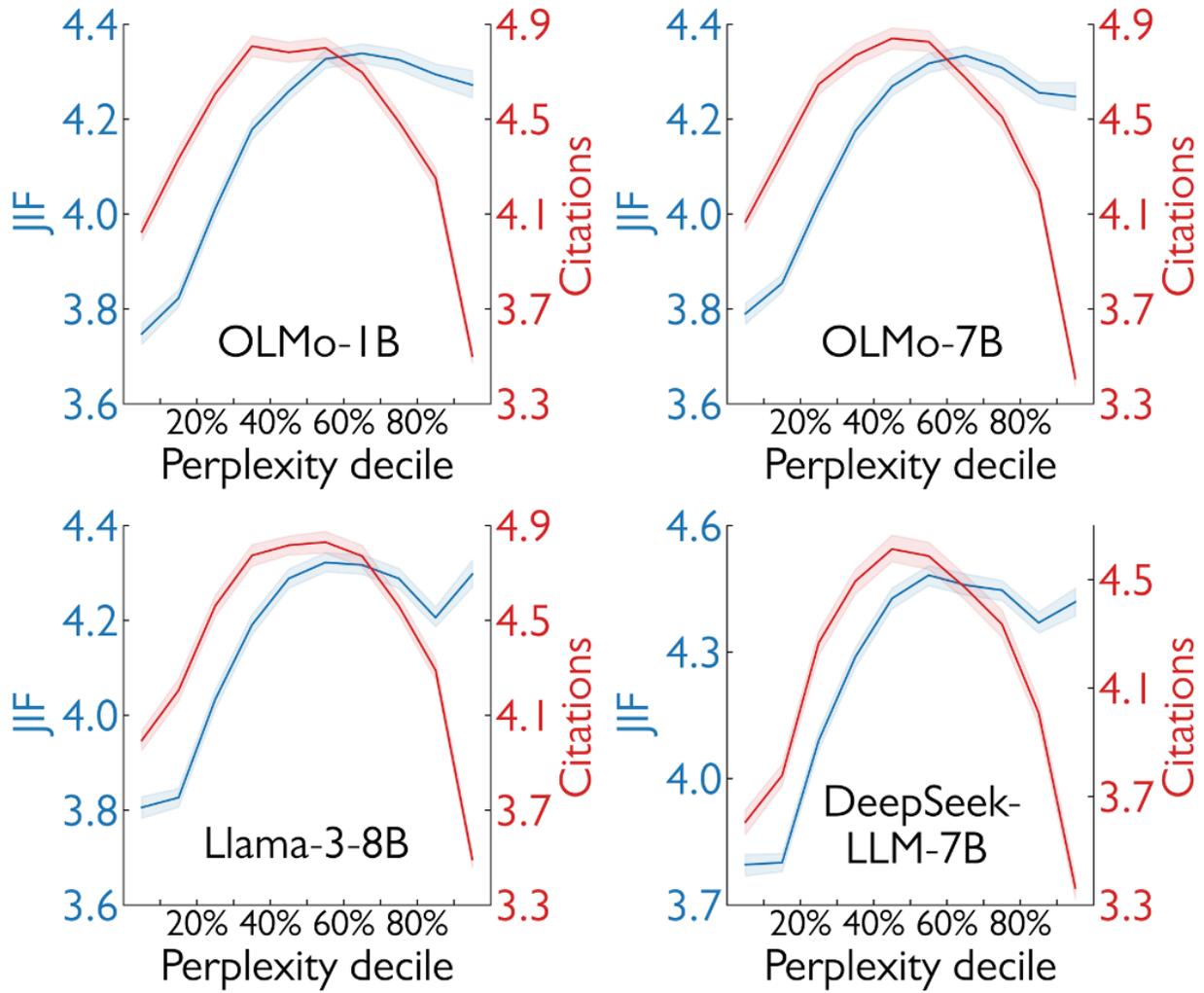

**Figure S3. Trends of average JIF and average citations across perplexity deciles calculated by four different models for papers in arts and humanities.**

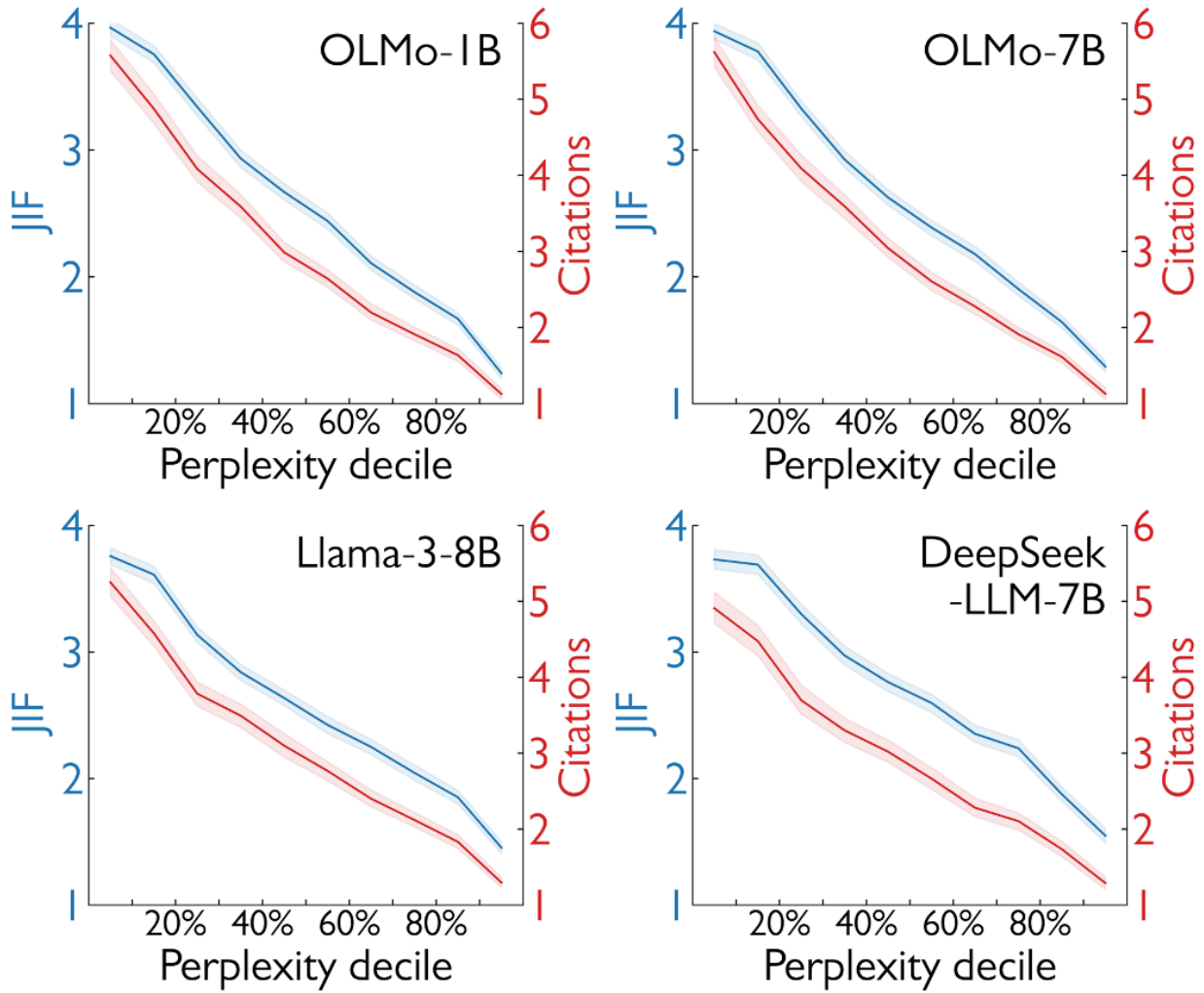

**Table S1. Pairwise Spearman correlations among perplexities calculated from four language models (OLMo-1B, OLMo-7B, Llama-3-8B, DeepSeek-LLM-7B) based on the WOS papers in natural and social sciences.**

|  | OLMo-1B | OLMo-7B | Llama-3-8B | DeepSeek-LLM-7B |
|---|---|---|---|---|
| **OLMo-1B** | 1 | | | |
| **OLMo-7B** | 0.98*** (n=1,784,196) | 1 | | |
| **Llama-3-8B** | 0.93*** (n=1,784,196) | 0.96*** (n=1,784,196) | 1 | |
| **DeepSeek-LLM-7B** | 0.96*** (n=1,355,619) | 0.97*** (n=1,355,619) | 0.96*** (n=1,355,619) | 1 |

*** $p < 0.001$, ** $p < 0.01$, * $p < 0.05$

**Table S2. Skewness of perplexity distributions across models and datasets**

| Dataset/model | Skewness | z | N |
|---|---|---|---|
| WOS/OLMo-1B | 3.11*** | 1701.20 | 1,784,017 |
| WOS/OLMo-7B | 3.12*** | 1702.54 | 1,784,017 |
| WOS/Llama-3-8B | 4.02*** | 2191.01 | 1,784,017 |
| WOS/DeepSeek-LLM-7B | 4.50*** | 2142.88 | 1,357,396 |
| OpenReview/OLMo-1B | 1.22*** | 44.58 | 8,005 |
| OpenReview/OLMo-7B | 1.28*** | 46.61 | 8,005 |
| OpenReview/Llama-3-8B | 1.40*** | 51.07 | 8,005 |
| OpenReview/DeepSeek-LLM-7B | 1.47*** | 53.91 | 8,005 |
| Semantic Scholar/OLMo-1B | 1.22*** | 51.87 | 10,766 |
| Semantic Scholar/OLMo-7B | 1.24*** | 52.48 | 10,766 |
| Semantic Scholar/Llama-3-8B | 1.42*** | 60.25 | 10,766 |
| Semantic Scholar/DeepSeek-LLM-7B | 1.42*** | 60.12 | 10,766 |
| Acceptance delay/GPT-2 | 2.09*** | 464.98 | 297,071 |

Materials and Methods M1.1 (WOS), M1.2 (OpenReview), M1.3 (Semantic Scholar), M1.4 (Acceptance delay). For WOS, only papers in the natural and social sciences are included.
*** $p < 0.001$, ** $p < 0.01$, * $p < 0.05$

**Table S3. Welch's t-test Results for Perplexity: Breakthrough vs. WOS Papers**

| Model | Breakthrough papers | WOS papers | t | Cohen's d |
|---|---|---|---|---|
| OLMo-1B | 18.70 ± 6.20 (n=33) | 13.62 ± 7.48 (n=1,784,177) | 4.70*** | 0.68 |
| OLMo-7B | 14.31 ± 4.34 (n=33) | 10.58 ± 5.64 (n=1,784,177) | 4.93*** | 0.66 |
| Llama-3-8B | 11.79 ± 3.70 (n=33) | 9.19 ± 5.85 (n=1,784,177) | 4.04*** | 0.45 |
| DeepSeek-LLM-7B | 13.30 ± 4.68 (n=33) | 10.42 ± 6.54 (n=1,358,710) | 3.54** | 0.44 |

*** $p < 0.001$, ** $p < 0.01$, * $p < 0.05$

**Table S4. Welch's t-test results for perplexity between papers nominated by scholars as surprising and unsurprising.**

| Model | Surprising | Unsurprising | t | Cohen's d |
|---|---|---|---|---|
| OLMo-1B | 21.80 ± 7.56 (n=33) | 13.12 ± 4.75 (n=20) | 5.14*** | 1.31 |
| OLMo-7B | 16.22 ± 5.07 (n=33) | 10.38 ± 3.68 (n=20) | 4.85*** | 1.27 |
| Llama-3-8B | 13.20 ± 3.95 (n=33) | 9.63 ± 3.72 (n=20) | 3.31** | 0.92 |
| DeepSeek-LLM-7B | 15.59 ± 4.76 (n=33) | 10.91 ± 3.91 (n=20) | 3.89*** | 1.05 |

\*\*\* $p < 0.001$, \*\* $p < 0.01$, \* $p < 0.05$

**Table S5. Distinguishing words and their ratios (perplexity was calculated using the OLMo-1B model).** We select titles and abstracts from 1,784,215 WOS journal articles published after the knowledge cutoff date of OLMo-1B and then assign them into two groups, including a perplexing group (perplexity in top 50%) and a non-perplexing group (perplexity in bottom 50%). For words observed in both groups (125,086 words for titles and 433,694 words for abstracts), we calculate *r*, the ratio of their frequency in perplexing to that in non-perplexing groups. We present some popular words (with high probability in both groups) whose ratios deviate significantly from 1.

| Distinguishing words | Title | Abstract |
|---|---|---|
| | Relative ratio *r* | |
| *Surprising (r>1) or unsurprising (r<1) Verbs* | | |
| create | **2.98** | 1.99 |
| discover | **2.25** | 1.21 |
| find | **1.78** | 0.80 |
| introduce | **1.69** | 2.04 |
| challenge | **1.44** | 1.61 |
| change | **1.40** | 1.21 |
| improve | **0.80** | 0.77 |
| modify | **0.67** | 0.83 |
| confirm | **0.56** | 0.67 |
| promote | **0.49** | 0.67 |
| compare | **0.40** | 0.50 |
| follow | **0.30** | 0.44 |
| *Surprising (r>1) or unsurprising (r<1) nouns* | | |
| paradigm | **2.17** | 2.45 |
| innovation | **1.53** | 1.41 |
| tool | **1.48** | 1.27 |
| exploration | **1.43** | 1.63 |
| device | **1.34** | 1.43 |
| technique | **1.21** | 1.30 |
| synthesis | **0.83** | 0.90 |
| survey | **0.67** | 0.73 |
| examination | **0.67** | 0.48 |
| validation | **0.57** | 0.78 |
| review | **0.56** | 0.63 |
| relationship | **0.51** | 0.66 |
| *Surprising (r>1) or unsurprising (r<1) adjectives* | | |
| extraordinary | **2.91** | 2.34 |
| unconventional | **2.22** | 2.45 |
| unprecedented | **2.01** | 2.17 |
| disruptive | **1.97** | 1.73 |
| unexpected | **1.57** | 1.84 |
| new | **1.37** | 1.40 |
| typical | **0.82** | 1.46 |
| normal | **0.67** | 0.54 |
| conventional | **0.51** | 1.08 |
| related | **0.39** | 0.55 |
| usual | **0.36** | 0.61 |
| old | **0.36** | 0.34 |

**Table S6. Distinguishing words and their ratios (perplexity was calculated using the OLMo-7B model).** As in Table S5, but we use OLMo-7B instead of OLMo-1B to compute the perplexity of 1,784,215 paper abstracts.

| Distinguishing words | Title | Abstract |
|---|---|---|
| | \multicolumn{2}{c}{Relative ratio *r*} | |
| \multicolumn{3}{c}{*Surprising (r>1) or unsurprising (r<1) verbs*} | | |
| create | **3.50** | 2.16 |
| discover | **2.02** | 1.21 |
| find | **1.87** | 0.80 |
| introduce | **1.60** | 2.02 |
| challenge | **1.60** | 1.65 |
| change | **1.52** | 1.25 |
| improve | **0.82** | 0.79 |
| modify | **0.61** | 0.79 |
| confirm | **0.51** | 0.65 |
| promote | **0.49** | 0.69 |
| compare | **0.42** | 0.50 |
| follow | **0.31** | 0.45 |
| \multicolumn{3}{c}{*Surprising (r>1) or unsurprising (r<1) nouns*} | | |
| paradigm | **2.38** | 2.62 |
| innovation | **1.71** | 1.60 |
| tool | **1.64** | 1.39 |
| exploration | **1.55** | 1.66 |
| device | **1.34** | 1.40 |
| technique | **1.27** | 1.35 |
| synthesis | **0.75** | 0.86 |
| survey | **0.74** | 0.81 |
| examination | **0.71** | 0.50 |
| review | **0.67** | 0.76 |
| validation | **0.57** | 0.80 |
| relationship | **0.53** | 0.70 |
| *Surprising (r>1) or unsurprising (r<1) adjectives* | | |
| extraordinary | **2.45** | 2.22 |
| unconventional | **2.32** | 2.53 |
| disruptive | **2.21** | 2.00 |
| unprecedented | **1.77** | 2.31 |
| unexpected | **1.45** | 1.81 |
| new | **1.29** | 1.41 |
| typical | **0.80** | 1.40 |
| normal | **0.66** | 0.53 |
| conventional | **0.52** | 1.09 |
| related | **0.40** | 0.58 |
| old | **0.38** | 0.36 |
| usual | **0.36** | 0.60 |

**Table S7. Distinguishing words and their ratios (perplexity was calculated using the Llama-3-B model).** As in Table S5, but we use Llama-3-8B instead of OLMo-1B to compute the perplexity of 1,784,215 paper abstracts.

| Distinguishing words | Title | Abstract |
|---|---|---|
| | Relative ratio *r* | |
| *Surprising (r>1) or unsurprising (r<1) verbs* | | |
| create | **4.03** | 2.37 |
| find | **1.86** | 0.78 |
| discover | **1.79** | 1.12 |
| challenge | **1.73** | 1.75 |
| change | **1.63** | 1.27 |
| introduce | **1.56** | 2.16 |
| improve | **0.89** | 0.88 |
| modify | **0.54** | 0.73 |
| promote | **0.46** | 0.72 |
| compare | **0.43** | 0.51 |
| confirm | **0.42** | 0.58 |
| follow | **0.29** | 0.43 |
| *Surprising (r>1) or unsurprising (r<1) nouns* | | |
| innovation | **2.74** | 2.49 |
| paradigm | **2.69** | 2.94 |
| tool | **1.87** | 1.55 |
| exploration | **1.81** | 1.84 |
| device | **1.38** | 1.43 |
| technique | **1.34** | 1.42 |
| survey | **0.87** | 0.99 |
| examination | **0.82** | 0.47 |
| review | **0.78** | 0.89 |
| validation | **0.60** | 0.84 |
| relationship | **0.60** | 0.82 |
| synthesis | **0.55** | 0.70 |
| *Surprising (r>1) or unsurprising (r<1) adjectives* | | |
| disruptive | **2.74** | 2.56 |
| unconventional | **2.09** | 2.26 |
| extraordinary | **1.94** | 2.13 |
| unprecedented | **1.43** | 2.23 |
| new | **1.18** | 1.40 |
| unexpected | **1.13** | 1.76 |
| typical | **0.83** | 1.34 |
| normal | **0.66** | 0.50 |
| conventional | **0.55** | 1.12 |
| related | **0.43** | 0.63 |
| old | **0.43** | 0.37 |
| usual | **0.39** | 0.65 |

**Table S8. Distinguishing words and their ratios (perplexity was calculated using the DeepSeek-LLM-7B model).** As in Table S5, but we use DeepSeek-LLM-7B instead of OLMo-1B to compute the perplexity of 1,358,755 paper abstracts, all published after the knowledge cutoff date of DeepSeek-LLM-7B.

| Distinguishing words | Title | Abstract |
|---|---|---|
| | \multicolumn{2}{c}{Relative ratio $r$} |
| \multicolumn{3}{c}{*Surprising ($r>1$) or unsurprising ($r<1$) verbs*} |
| create | **3.44** | 2.06 |
| discover | **1.70** | 1.08 |
| find | **1.66** | 0.77 |
| introduce | **1.65** | 2.32 |
| challenge | **1.50** | 1.62 |
| change | **1.44** | 1.25 |
| improve | **0.83** | 0.82 |
| modify | **0.61** | 0.80 |
| promote | **0.43** | 0.66 |
| confirm | **0.38** | 0.60 |
| compare | **0.37** | 0.50 |
| follow | **0.24** | 0.39 |
| \multicolumn{3}{c}{*Surprising ($r>1$) or unsurprising ($r<1$) nouns*} |
| paradigm | **2.18** | 2.77 |
| innovation | **1.98** | 1.84 |
| tool | **1.57** | 1.35 |
| exploration | **1.57** | 1.70 |
| device | **1.37** | 1.45 |
| technique | **1.29** | 1.39 |
| synthesis | **0.69** | 0.79 |
| examination | **0.67** | 0.41 |
| survey | **0.65** | 0.74 |
| review | **0.62** | 0.70 |
| validation | **0.55** | 0.75 |
| relationship | **0.54** | 0.74 |
| *Surprising (r>1) or unsurprising (r<1) adjectives* | | |
| unconventional | **2.64** | 2.70 |
| extraordinary | **2.54** | 2.16 |
| disruptive | **2.33** | 2.17 |
| new | **1.32** | 1.43 |
| unprecedented | **1.29** | 2.05 |
| unexpected | **1.22** | 1.71 |
| typical | **0.81** | 1.45 |
| normal | **0.75** | 0.51 |
| conventional | **0.50** | 1.14 |
| usual | **0.38** | 0.60 |
| related | **0.38** | 0.54 |
| old | **0.34** | 0.30 |

**Table S9. Chi-square test results for perplexing and non-perplexing term frequencies.**

| Model | Term type | N (High PPL group) | N (Low PPL group) | $\chi^2$ | df |
|---|---|---|---|---|---|
| OLMo-1B | Perplexing | 51,994 | 36,610 | 5354.25 *** | 1 |
| | Non-perplexing | 82,706 | 143,596 | 8926.06 *** | 1 |
| OLMo-7B | Perplexing | 52,076 | 36,299 | 5759.85 *** | 1 |
| | Non-perplexing | 85,232 | 140,827 | 6605.22 *** | 1 |
| Llama-3-8B | Perplexing | 53,359 | 36,313 | 6820.58 *** | 1 |
| | Non-perplexing | 87,453 | 139,207 | 4742.24 *** | 1 |
| DeepSeek-LLM-7B | Perplexing | 41,202 | 28,473 | 4866.08 *** | 1 |
| | Non-perplexing | 63,928 | 112,471 | 6756.78 *** | 1 |

**Standardized residual**

| Model | | High perplexity group | Low perplexity group |
|---|---|---|---|
| OLMo-1B | Perplexing terms | 72.318 | -62.512 |
| | Non-perplexing terms | -45.238 | 39.104 |
| OLMo-7B | Perplexing terms | 68.640 | -60.434 |
| | Non-perplexing terms | -42.917 | 37.786 |
| Llama-3-8B | Perplexing terms | 67.282 | -60.264 |
| | Non-perplexing terms | -42.320 | 37.905 |
| DeepSeek-LLM-7B | Perplexing terms | 66.277 | -57.240 |
| | Non-perplexing terms | -41.653 | 35.974 |

*** $p < 0.001$, ** $p < 0.01$, * $p < 0.05$

**Table S10. Logistic regression analysis of logged perplexity (calculated by GPT-2) predicting extremely long or short review time.**

| Threshold | OR | 95% CI | N |
|---|---|---|---|
| Top 1% | 1.23*** | [1.15, 1.33] | 297,101 |
| Top 3% | 1.12*** | [1.08, 1.17] | 297,101 |
| Bottom 1% | 1.92*** | [1.80, 2.06] | 297,101 |
| Bottom 3% | 1.83*** | [1.76, 1.91] | 297,101 |

*** $p < 0.001$, ** $p < 0.01$, * $p < 0.05$

**Table S11. Welch's t-test results for intra-rating disparity between high (top 20%) and low (bottom 20%) perplexity groups.**

| Model | High perplexity group | Low perplexity group | t | Cohen's d |
|---|---|---|---|---|
| OLMo-1B | 2.22 ± 1.33 (n=1,513) | 2.12±1.32 (n=1,548) | 2.23* | 0.08 |
| OLMo-7B | 2.23 ± 1.35 (n=1,513) | 2.10 ± 1.31 (n=1,552) | 2.76** | 0.10 |
| Llama-3-8B | 2.17 ± 1.34 (n=1,511) | 2.13±1.31 (n=1,552) | 0.77 | 0.03 |
| DeepSeek-LLM-7B | 2.17 ± 1.34 (n=1,514) | 2.12 ± 1.28 (n=1,551) | 1.12 | 0.04 |

*** $p < 0.001$, ** $p < 0.01$, * $p < 0.05$

**Table S12. Pearson correlations between logged perplexity and review confidence.**

| Model | r | 95% CI | N |
|---|---|---|---|
| OLMo-1B | -0.0309** | [-0.0528, -0.0090] | 8,005 |
| OLMo-7B | -0.0225* | [-0.0444, -0.0006] | 8,005 |
| Llama-3-8B | -0.0024 | [-0.0243, 0.0195] | 8,005 |
| DeepSeek-LLM-7B | -0.0501*** | [-0.0719, -0.0282] | 8,005 |

*** $p < 0.001$, ** $p < 0.01$, * $p < 0.05$

**Table S13. Logistic regression analysis of logged perplexity predicting extremely low confidence score (bottom 5%).**

| Model | OR | 95% CI | $R^2$ | N |
|---|---|---|---|---|
| OLMo-1B | 1.54** | [1.14, 2.08] | 0.0020 | 8,005 |
| OLMo-7B | 1.55** | [1.14, 2.11] | 0.0020 | 8,005 |
| Llama-3-8B | 1.19 | [0.89, 1.60] | 0.0004 | 8,005 |
| DeepSeek-LLM-7B | 1.82*** | [1.35, 2.45] | 0.0040 | 8,005 |

*** $p < 0.001$, ** $p < 0.01$, * $p < 0.05$

**Table S14. Chi-square test results for frequencies of uncertainty-related words in review comments.**

| Model | N (High PPL group) | N (Low PPL group) | $\chi^2$ | df |
|---|---|---|---|---|
| OLMo-1B | 2,024 | 1,621 | 14.73*** | 1 |
| OLMo-7B | 1,962 | 1,535 | 21.49*** | 1 |
| Llama-3-8B | 2,012 | 1,634 | 14.31*** | 1 |
| DeepSeek-LLM-7B | 1,949 | 1,697 | 12.56*** | 1 |

*** $p < 0.001$, ** $p < 0.01$, * $p < 0.05$

**Table S15. White's test for heteroskedasticity in review rating and confidence across models.**

| Model | Review rating | | Review confidence | |
|---|---|---|---|---|
| | White $\chi^2$ | N | White $\chi^2$ | N |
| OLMo-1B | 4.41 | 7,880 | 6.68* | 7,880 |
| OLMo-7B | 7.64* | 7,880 | 9.34** | 7,880 |
| Llama-3-8B | 29.94*** | 7,880 | 14.64*** | 7,880 |
| DeepSeek-LLM-7B | 3.55 | 7,880 | 8.88* | 7,880 |

*** $p < 0.001$, ** $p < 0.01$, * $p < 0.05$

**Table S16. Ordinary least squares (OLS) regression of variance in ratings.** OLS regression was performed to examine the relationship between perplexity and the variance in review ratings. For each model, papers were binned into 20 quantiles based on their perplexity scores, and the variance of review ratings was computed within each quantile. The quantile variable was then scaled to a 0-1 range to serve as the independent variable, and the log-transformed variance of review ratings was used as the dependent variable. Model identity was included as a categorical covariate to account for fixed effects across different language models.

| Variable | Coef | SE | 95% CI | t |
| --- | --- | --- | --- | --- |
| Intercept | 0.5585*** | 0.020 | [0.518, 0.599] | 27.787 |
| Perplexity bin (scaled) | 0.0407 | 0.026 | [-0.010, 0.092] | 1.591 |

$R^2 = 0.034$, N = 80, *** $p < 0.001$, ** $p < 0.01$, * $p < 0.05$

**Table S17. OLS regression of variance in confidences.** As in Table S16, but for confidences rather than ratings.

| Variable | Coef | SE | 95% CI | t |
| --- | --- | --- | --- | --- |
| Intercept | -1.3625*** | 0.025 | [-1.412, -1.313] | -55.042 |
| Perplexity bin (scaled) | 0.1028 | 0.031 | [0.040, 0.165] | 3.265 |

$R^2 = 0.128$, N = 80, *** $p < 0.001$, ** $p < 0.01$, * $p < 0.05$

**Table S18. White's test for heteroskedasticity and OLS regression of variance in review time.** Perplexity values were divided into 100 quantile-based bins, and the variance of review time was computed within each bin. The mean perplexity of each bin was used as the predictor, and the variance of review time as the dependent variable.

| White's test | | | | | |
|---|---|---|---|---|---|
| **White $\chi^2$** | | | N | | |
| 23.60*** | | | 297,101 | | |
| **OLS** | | | | | |
| **Variable** | **Coefficient** | **SE** | **95% CI** | | **t** |
| Intercept | 4016.14*** | 69.15 | [3878.91, 4153.37] | | 58.08 |
| Perplexity bin (scaled) | 30.06*** | 3.86 | [22.40, 37.72] | | 7.79 |
| $R^2 = 0.382$, N = 100, *** $p < 0.001$, ** $p < 0.01$, * $p < 0.05$ | | | | | |

**Table S19. White's test for heteroskedasticity of variance in JIF (natural and social science fields).**

| Field | OLMo-1B | OLMo-7B | Llama-3-8B | DeepSeek-LLM-7B |
|---|---|---|---|---|
| | | White $\chi^2$ | | |
| Agricultural Sciences | 1307.3758*** | 1254.8942*** | 1147.4032*** | 400.3573*** |
| Biology & Biochemistry | 16214.2364*** | 25529.5459*** | 0.1534 | 2.4874 |
| Chemistry | 14063.6271*** | 3363.9022*** | 3.5605 | 2.4942 |
| Clinical Medicine | 330.6456*** | 28.8871*** | 0.0018 | 0.0035 |
| Computer Science | 193.8316*** | 197.6265*** | 230.1344*** | 207.3678*** |
| Economics & Business | 353.9176*** | 367.3365*** | 2.1356 | 324.6689*** |
| Engineering | 329.6532*** | 242.7079*** | 0.2276 | 2.2938 |
| Environment/Ecology | 1949.4420*** | 675.3799*** | 0.6460 | 279.3735*** |
| Geosciences | 642.8009*** | 771.3352*** | 45.3033*** | 121.2981*** |
| Materials Science | 1005.1557*** | 911.7438*** | 559.0556*** | 434.5577*** |
| Mathematics | 2804.3507*** | 1069.5532*** | 173.0593*** | 57.1257*** |
| Multidisciplinary | 5352.7217*** | 3902.3893*** | 31.0779*** | 232.8232*** |
| Physics | 837.3387*** | 708.1987*** | 29.0838*** | 37.9033*** |
| Plant & Animal Science | 1413.7174*** | 1436.6054*** | 2339.9449*** | 1506.6711*** |
| Psychiatry/Psychology | 1337.0336*** | 1287.8482*** | 261.6221*** | 59.6931*** |
| Social Sciences, General | 1158.3847*** | 954.9401*** | 0.9260 | 9.5725** |

*** $p < 0.001$, ** $p < 0.01$, * $p < 0.05$

**Table S20. OLS regression of variance in JIF (natural and social science fields).** For each research field, data from four language models (OLMo-1B, OLMo-7B, Llama3-8B, and DeepSeek-LLM-7B) were included. Papers were grouped into 1,000 quantile-based bins according to their perplexity scores. Within each bin, the variance of JIF was computed, and the logged variance was used as the dependent variable. The independent variable was the bin index scaled from 0 to 1, representing increasing perplexity. Fixed effects for language models were included to control for systematic differences across models.

| Field | Coefficient | SE | 95% CI | $R^2$ | N |
|---|---|---|---|---|---|
| Agricultural Sciences | 0.8073*** | 0.017 | [0.774, 0.840] | 0.365 | 4,000 |
| Biology & Biochemistry | 1.5426*** | 0.031 | [1.481, 1.604] | 0.378 | 4,000 |
| Chemistry | 1.1563*** | 0.021 | [1.114, 1.198] | 0.422 | 4,000 |
| Clinical Medicine | 0.4539*** | 0.041 | [0.374, 0.534] | 0.032 | 4,000 |
| Computer Science | 0.4352*** | 0.014 | [0.408, 0.462] | 0.200 | 4,000 |
| Economics & Business | 0.4980*** | 0.014 | [0.470, 0.526] | 0.230 | 4,000 |
| Engineering | 0.1831*** | 0.012 | [0.160, 0.206] | 0.058 | 4,000 |
| Environment/Ecology | 0.5813*** | 0.013 | [0.555, 0.607] | 0.327 | 4,000 |
| Geosciences | 0.5174*** | 0.033 | [0.453, 0.582] | 0.059 | 4,000 |
| Materials Science | 0.5887*** | 0.017 | [0.556, 0.622] | 0.236 | 4,000 |
| Mathematics | -0.0426* | 0.019 | [-0.079, -0.006] | 0.003 | 4,000 |
| Multidisciplinary | 0.8226*** | 0.017 | [0.789, 0.856] | 0.368 | 4,000 |
| Physics | 0.4759*** | 0.017 | [0.443, 0.509] | 0.167 | 4,000 |
| Plant & Animal Science | 0.7451*** | 0.017 | [0.712, 0.778] | 0.326 | 4,000 |
| Psychiatry/Psychology | 0.6820*** | 0.034 | [0.615, 0.750] | 0.090 | 4,000 |
| Social Sciences, General | 0.5293*** | 0.018 | [0.495, 0.564] | 0.183 | 4,000 |

*** $p < 0.001$, ** $p < 0.01$, * $p < 0.05$

**Table S21. White's test for heteroskedasticity of variance in JIF (arts and humanities fields).**

| Field | OLMo-1B | OLMo-7B | Llama-3-8B | DeepSeek-LLM-7B |
|---|---|---|---|---|
| | | White $\chi^2$ | | |
| Arts & Humanities, Interdisciplinary | 303.1804*** | 235.4784*** | 17.1468*** | 103.4385*** |
| History & Archaeology | 8.9960* | 6.3332* | 0.1952 | 0.1782 |
| Literature & Language | 6411.2798*** | 7362.3508*** | 1449.3705*** | 21.6888*** |
| Philosophy & Religion | 516.9567*** | 246.0641*** | 67.8187*** | 90.3065*** |
| Visual & Performing Arts | 588.6024*** | 77.2430*** | 1091.5247*** | 26.2633*** |

*** $p < 0.001$, ** $p < 0.01$, * $p < 0.05$

**Table S22. OLS regression of variance in JIF (arts and humanities fields).** As in Table S20, but for 5 arts and humanities fields rather than natural and social science fields, with different LLMs as fixed effect. For analyses in arts and humanities fields, papers were grouped into 100 bins due to sample size limitations.

| Field | Coefficient | SE | 95% CI | $R^2$ | N |
|---|---|---|---|---|---|
| Arts & Humanities, Interdisciplinary | -0.6848*** | 0.031 | [-0.746, -0.624] | 0.552 | 400 |
| History & Archaeology | -0.0826* | 0.039 | [-0.159, -0.006] | 0.014 | 400 |
| Literature & Language | -0.3116*** | 0.033 | [-0.376, -0.247] | 0.188 | 400 |
| Philosophy & Religion | -1.3773*** | 0.075 | [-1.524, -1.231] | 0.465 | 400 |
| Visual & Performing Arts | 0.1759*** | 0.024 | [0.129, 0.223] | 0.127 | 400 |

*** $p < 0.001$, ** $p < 0.01$, * $p < 0.05$

**Table S23. Pearson correlations between logged perplexity and review rating.**

| Model | r | 95% CI | N |
| --- | --- | --- | --- |
| OLMo-1B | 0.07*** | [0.05, 0.10] | 7,880 |
| OLMo-7B | 0.07*** | [0.05, 0.09] | 7,880 |
| Llama-3-8B | 0.03** | [0.01, 0.05] | 7,880 |
| DeepSeek-LLM-7B | 0.06*** | [0.03, 0.08] | 7,880 |

*** $p < 0.001$, ** $p < 0.01$, * $p < 0.05$

**Table S24. Logistic regression analysis of logged perplexity predicting extremely high rating score (top 5%).**

| Model | OR | 95% CI | $R^2$ | N |
|---|---|---|---|---|
| OLMo-1B | 1.92*** | [1.42, 2.59] | 0.0046 | 8,005 |
| OLMo-7B | 1.93*** | [1.42, 2.62] | 0.0045 | 8,005 |
| Llama-3-8B | 2.10*** | [1.57, 2.82] | 0.0062 | 8,005 |
| DeepSeek-LLM-7B | 1.69** | [1.20, 2.38] | 0.0029 | 8,005 |

*** $p < 0.001$, ** $p < 0.01$, * $p < 0.05$

**Table S25. Logistic regression analysis of logged perplexity predicting extremely low rating score (bottom 5%).**

| Model | OR | 95% CI | $R^2$ | N |
|---|---|---|---|---|
| OLMo-1B | 0.56** | [0.39, 0.80] | 0.0033 | 8,005 |
| OLMo-7B | 0.65* | [0.46, 0.94] | 0.0017 | 8,005 |
| Llama-3-8B | 0.99 | [0.70, 1.39] | $1.55 \times 10^{-6}$ | 8,005 |
| DeepSeek-LLM-7B | 0.67* | [0.47, 0.95] | 0.0016 | 8,005 |

*** $p < 0.001$, ** $p < 0.01$, * $p < 0.05$

**Table S26. Mann-Whitney U test results comparing perplexities between award and non-award papers.**

| Model | Award paper | Non-award paper | U | Cohen's d |
|---|---|---|---|---|
| OLMo-1B | 19.54 ± 6.12 (n=127) | 18.02 ± 5.38 (n=10,639) | 784143.00 ** | 0.28 |
| OLMo-7B | 14.84 ± 4.43 (n=127) | 13.71 ± 4.02 (n=10,639) | 790016.00 ** | 0.28 |
| Llama-3-8B | 13.61 ± 4.99 (n=127) | 12.81 ± 3.90 (n=10,639) | 753687.50 * | 0.20 |
| DeepSeek-LLM-7B | 14.93 ± 5.46 (n=127) | 13.98 ± 4.25 (n=10,639) | 753536.00 * | 0.22 |

\*\*\* $p < 0.001$, \*\* $p < 0.01$, \* $p < 0.05$, † $p < 0.1$.

**Table S27. Logistic regression analysis of perplexity predicting award-winning papers.**

| Variable | Dependent variable: is award-winning papers | | | | | | | |
|---|---|---|---|---|---|---|---|---|
| | Models | | | | | | | |
| | OLMo-1B | | OLMo-7B | | Llama-3-8B | | DeepSeek-llm-7B | |
| | OR | 95% CI | OR | 95% CI | OR | 95% CI | OR | 95% CI |
| Log (PPL) | 2.63 ** | [1.44, 4.79] | 2.76 ** | [1.50, 5.07] | 2.05 * | [1.13, 3.72] | 2.09 * | [1.15, 3.79] |
| $R^2$ | 0.0070 | | 0.0075 | | 0.0040 | | 0.0042 | |
| $N$ | 10,766 | | 10,766 | | 10,766 | | 10,766 | |
| | Models | | | | | | | |
| | OLMo-1B | | OLMo-7B | | Llama-3-8B | | DeepSeek-llm-7B | |
| | OR | 95% CI | OR | 95% CI | OR | 95% CI | OR | 95% CI |
| Log (Perplexity) | 3.08 ** | [1.39, 6.83] | 3.39 ** | [1.50, 7.68] | 2.60 ** | [1.16, 5.82] | 2.91 ** | [1.31, 6.49] |
| Abstract length | Controlled | | | | | | | |
| Venue | Fixed effects | | | | | | | |
| $R^2$ | 0.0721 | | 0.0731 | | 0.0705 | | 0.0708 | |
| $N$ | 10,129 | | 10,129 | | 10,129 | | 10,129 | |

*** $p < 0.001$, ** $p < 0.01$, * $p < 0.05$

**Table S28. Logistic regression of funder on logged perplexity.**

| Funder | Model | OR | 95% CI | $R^2$ | N |
|---|---|---|---|---|---|
| DARPA | OLMo-1B | 3.63*** | [3.26, 4.05] | 0.02411 | 1,784,196 |
|  | OLMo-7B | 3.37*** | [3.02, 3.77] | 0.02051 | 1,784,196 |
|  | Llama-3-8B | 2.81*** | [2.55, 3.11] | 0.01691 | 1,784,196 |
|  | DeepSeek-LLM-7B | 2.21*** | [2.03, 2.41] | 0.01686 | 1,358,755 |
| AFOSR | OLMo-1B | 4.06*** | [3.77, 4.37] | 0.03142 | 1,784,196 |
|  | OLMo-7B | 3.66*** | [3.40, 3.96] | 0.02543 | 1,784,196 |
|  | Llama-3-8B | 2.60*** | [2.41, 2.80] | 0.01430 | 1,784,196 |
|  | DeepSeek-LLM-7B | 2.58*** | [2.39, 2.77] | 0.02209 | 1,358,755 |
| ONR | OLMo-1B | 3.61*** | [3.34, 3.91] | 0.02530 | 1,784,196 |
|  | OLMo-7B | 3.37*** | [3.11, 3.66] | 0.02153 | 1,784,196 |
|  | Llama-3-8B | 2.64*** | [2.45, 2.85] | 0.01485 | 1,784,196 |
|  | DeepSeek-LLM-7B | 2.78*** | [2.57, 3.01] | 0.02233 | 1,358,755 |
| NSF | OLMo-1B | 2.76*** | [2.70, 2.81] | 0.02221 | 1,784,196 |
|  | OLMo-7B | 2.44*** | [2.39, 2.49] | 0.01597 | 1,784,196 |
|  | Llama-3-8B | 1.85*** | [1.82, 1.89] | 0.00797 | 1,784,196 |
|  | DeepSeek-LLM-7B | 2.23*** | [2.18, 2.28] | 0.01517 | 1,358,755 |
| NIH | OLMo-1B | 0.57*** | [0.56, 0.58] | 0.00600 | 1,784,196 |
|  | OLMo-7B | 0.54*** | [0.53, 0.55] | 0.00670 | 1,784,196 |
|  | Llama-3-8B | 0.46*** | [0.45, 0.47] | 0.01082 | 1,784,196 |
|  | DeepSeek-LLM-7B | 0.46*** | [0.45, 0.47] | 0.01224 | 1,358,755 |
| ERC | OLMo-1B | 3.04*** | [2.94, 3.15] | 0.02252 | 1,784,196 |
|  | OLMo-7B | 2.69*** | [2.59, 2.79] | 0.01653 | 1,784,196 |
|  | Llama-3-8B | 1.92*** | [1.86, 1.99] | 0.00742 | 1,784,196 |
| | DeepSeek-LLM-7B | 2.23*** | [2.15, 2.32] | 0.01373 | 1,358,755 |
| DFG | OLMo-1B | 2.65*** | [2.57, 2.73] | 0.01766 | 1,784,196 |
| | OLMo-7B | 2.33*** | [2.25, 2.40] | 0.01231 | 1,784,196 |
| | Llama-3-8B | 1.63*** | [1.58, 1.69] | 0.00424 | 1,784,196 |
| | DeepSeek-LLM-7B | 2.00*** | [1.94, 2.07] | 0.01033 | 1,358,755 |
| EC | OLMo-1B | 2.54*** | [2.48, 2.61] | 0.01711 | 1,784,196 |
| | OLMo-7B | 2.46*** | [2.40, 2.53] | 0.01495 | 1,784,196 |
| | Llama-3-8B | 2.07*** | [2.02, 2.13] | 0.01036 | 1,784,196 |
| | DeepSeek-LLM-7B | 2.22*** | [2.16, 2.28] | 0.01422 | 1,358,755 |
| NSERC | OLMo-1B | 2.11*** | [2.02, 2.19] | 0.00928 | 1,784,196 |
| | OLMo-7B | 1.94*** | [1.86, 2.02] | 0.00678 | 1,784,196 |
| | Llama-3-8B | 1.55*** | [1.49, 1.61] | 0.00307 | 1,784,196 |
| | DeepSeek-LLM-7B | 1.79*** | [1.72, 1.87] | 0.00691 | 1,358,755 |
| ARC | OLMo-1B | 1.96*** | [1.87, 2.05] | 0.00718 | 1,784,196 |
| | OLMo-7B | 1.82*** | [1.74, 1.91] | 0.00535 | 1,784,196 |
| | Llama-3-8B | 1.61*** | [1.54, 1.68] | 0.00350 | 1,784,196 |
| | DeepSeek-LLM-7B | 1.70*** | [1.63, 1.78] | 0.00556 | 1,358,755 |
| NSFC | OLMo-1B | 1.01* | [1.00, 1.02] | $3 \times 10^{-6}$ | 1,784,196 |
| | OLMo-7B | 0.89*** | [0.88, 0.90] | 0.00035 | 1,784,196 |
| | Llama-3-8B | 0.88*** | [0.88, 0.89] | 0.00041 | 1,784,196 |
| | DeepSeek-LLM-7B | 1.12*** | [1.11, 1.14] | 0.00043 | 1,358,755 |
| NSSFC | OLMo-1B | 1.08 | [1.00, 1.18] | $8 \times 10^{-5}$ | 1,784,196 |
| | OLMo-7B | 1.10* | [1.01, 1.20] | 0.00011 | 1,784,196 |
| | Llama-3-8B | 1.77*** | [1.63, 1.92] | 0.00438 | 1,784,196 |
| | DeepSeek-LLM-7B | 1.48*** | [1.37, 1.61] | 0.00244 | 1,358,755 |
| JSPS | OLMo-1B | 1.04* | [1.01, 1.07] | $3\times10^{-5}$ | 1,784,196 |
| | OLMo-7B | 0.92*** | [0.90, 0.96] | 0.00010 | 1,784,196 |
| | Llama-3-8B | 0.70*** | [0.67, 0.72] | 0.00214 | 1,784,196 |
| | DeepSeek-LLM-7B | 0.92*** | [0.89, 0.96] | 0.00012 | 1,358,755 |
| ARF | OLMo-1B | 0.86*** | [0.84, 0.89] | 0.00037 | 1,784,196 |
| | OLMo-7B | 0.85*** | [0.82, 0.87] | 0.00044 | 1,784,196 |
| | Llama-3-8B | 0.82*** | [0.80, 0.85] | 0.00065 | 1,784,196 |
| | DeepSeek-LLM-7B | 0.90*** | [0.87, 0.93] | 0.00020 | 1,358,755 |

*** $p < 0.001$, ** $p < 0.01$, * $p < 0.05$

**Table S29. Logistic regression of top 5% IF on logged perplexity (natural and social sciences).**

| Variable | OLMo-1B | | OLMo-7B | | Llama-3-8B | | DeepSeek-LLM-7B | |
|---|---|---|---|---|---|---|---|---|
| | OR | 95% CI | OR | 95% CI | OR | 95% CI | OR | 95% CI |
| Log (PPL) | 1.9631 *** | [1.9337, 1.9929] | 1.8850 *** | [1.8558, 1.9145] | 1.7234 *** | [1.6975, 1.7496] | 1.7595 *** | [1.7305, 1.7889] |
| N | 1,774,065 | | 1,774,065 | | 1,774,065 | | 1,351,531 | |
| $R^2$ | 0.011 | | 0.009 | | 0.007 | | 0.008 | |

*** $p < 0.001$, ** $p < 0.01$, * $p < 0.05$

**Table S30. Logistic regression of top 5% IF on logged perplexity in each field (natural and social science fields), with different models as fixed effect.**

| Field | OR | 95% CI | $R^2$ | N |
|---|---|---|---|---|
| Agricultural Sciences | 4.5028*** | [4.229, 4.794] | 0.038 | 134,943 |
| Biology & Biochemistry | 3.6392*** | [3.577, 3.702] | 0.034 | 1,522,864 |
| Chemistry | 2.6178*** | [2.575, 2.661] | 0.018 | 1,572,668 |
| Clinical Medicine | 1.7110*** | [1.688, 1.735] | 0.007 | 2,087,574 |
| Computer Science | 1.5842*** | [1.535, 1.635] | 0.004 | 483,264 |
| Economics & Business | 1.2021*** | [1.168, 1.238] | 0.001 | 612,888 |
| Engineering | 0.9410*** | [0.923, 0.959] | $6 \times 10^{-5}$ | 1,631,905 |
| Environment/Ecology | 1.8535*** | [1.801, 1.907] | 0.007 | 637,572 |
| Geosciences | 1.2278*** | [1.176, 1.282] | 0.001 | 330,305 |
| Materials Science | 1.6302*** | [1.594, 1.667] | 0.004 | 1,098,531 |
| Mathematics | 0.9479** | [0.916, 0.980] | $5 \times 10^{-5}$ | 453,874 |
| Multidisciplinary | 1.7793*** | [1.754, 1.805] | 0.007 | 2,153,160 |
| Physics | 1.2422*** | [1.220, 1.265] | 0.001 | 1,615,309 |
| Plant & Animal Science | 2.8252*** | [2.729, 2.924] | 0.020 | 430,804 |
| Psychiatry/Psychology | 1.7501*** | [1.714, 1.787] | 0.007 | 864,617 |
| Social Sciences, General | 1.1234*** | [1.098, 1.149] | $5 \times 10^{-4}$ | 803,197 |

*** $p < 0.001$, ** $p < 0.01$, * $p < 0.05$

**Table S31. Logistic regression of bottom 5% IF on logged perplexity (natural and social sciences).**

| Variable | OLMo-1B | | OLMo-7B | | Llama-3-8B | | DeepSeek-LLM-7B | |
|---|---|---|---|---|---|---|---|---|
| | OR | 95% CI | OR | 95% CI | OR | 95% CI | OR | 95% CI |
| Log (PPL) | 1.9983 *** | [1.9683, 2.0287] | 2.0539 *** | [2.0222, 2.0861] | 1.8453 *** | [1.8176, 1.8734] | 1.6481 *** | [1.6210, 1.6756] |
| N | 1,774,065 | | 1,774,065 | | 1,774,065 | | 1,351,531 | |
| $R^2$ | 0.011 | | 0.011 | | 0.009 | | 0.006 | |

*** $p < 0.001$, ** $p < 0.01$, * $p < 0.05$

**Table S32. Logistic regression of bottom 5% IF on logged perplexity in each field (natural and social science fields), with different models as fixed effect.**

| Field | OR | 95% CI | $R^2$ | N |
|---|---|---|---|---|
| Agricultural Sciences | 1.0164 | [0.950, 1.088] | $1\times10^{-5}$ | 134,943 |
| Biology & Biochemistry | 1.0361*** | [1.017, 1.055] | $3\times10^{-5}$ | 1,522,864 |
| Chemistry | 0.8437*** | [0.828, 0.860] | 0.001 | 1,572,668 |
| Clinical Medicine | 1.7776*** | [1.753, 1.802] | 0.008 | 2,087,574 |
| Computer Science | 1.9225*** | [1.861, 1.986] | 0.008 | 483,264 |
| Economics & Business | 2.7885*** | [2.709, 2.870] | 0.020 | 612,888 |
| Engineering | 1.5928*** | [1.564, 1.622] | 0.004 | 1,631,905 |
| Environment/Ecology | 1.2476*** | [1.212, 1.284] | 0.001 | 637,572 |
| Geosciences | 1.1027*** | [1.056, 1.152] | $1\times10^{-4}$ | 330,305 |
| Materials Science | 0.9984 | [0.975, 1.022] | $10\times10^{-7}$ | 1,098,531 |
| Mathematics | 1.5452*** | [1.494, 1.598] | 0.003 | 453,874 |
| Multidisciplinary | 3.4769*** | [3.426, 3.529] | 0.031 | 2,153,160 |
| Physics | 1.1318*** | [1.111, 1.153] | $3\times10^{-4}$ | 1,615,309 |
| Plant & Animal Science | 1.6245*** | [1.570, 1.681] | 0.004 | 430,804 |
| Psychiatry/Psychology | 3.2533*** | [3.184, 3.324] | 0.033 | 864,617 |
| Social Sciences, General | 3.7729*** | [3.685, 3.863] | 0.040 | 803,197 |

*** $p < 0.001$, ** $p < 0.01$, * $p < 0.05$

**Table S33. Logistic regression analysis of perplexity predicting high and low impact factor publications, with abstract length, field, and publication date controlled (natural and social sciences).**

| Variable | Dependent variable: is in top 5% JIF | | | | | | | |
|---|---|---|---|---|---|---|---|---|
| | Models | | | | | | | |
| | OLMo-1B | | OLMo-7B | | Llama-3-8B | | DeepSeek-LLM-7B | |
| | OR | 95% CI | OR | 95% CI | OR | 95% CI | OR | 95% CI |
| Log (Perplexity) | 2.03 *** | [1.99, 2.07] | 1.92 *** | [1.88, 1.96] | 1.84 *** | [1.80, 1.88] | 1.85 *** | [1.81, 1.90] |
| Abstract length | Controlled | | | | | | | |
| Field | Fixed effects | | | | | | | |
| Pub. date | Fixed effects | | | | | | | |
| $R^2$ | 0.1008 | | 0.0993 | | 0.0982 | | 0.1069 | |
| N | 1,763,932 | | 1,763,932 | | 1,763,932 | | 1,342,012 | |
| Variable | Dependent variable: is in bottom 5% JIF | | | | | | | |
| | Models | | | | | | | |
| | OLMo-1B | | OLMo-7B | | Llama-3-8B | | DeepSeek-LLM-7B | |
| | OR | 95% CI | OR | 95% CI | OR | 95% CI | OR | 95% CI |
| Log (Perplexity) | 1.36 *** | [1.34, 1.39] | 1.37 *** | [1.34, 1.39] | 1.17 *** | [1.15, 1.20] | 1.12 *** | [1.10, 1.15] |
| Abstract length | Controlled | | | | | | | |
| Field | Fixed effects | | | | | | | |
| Pub. date | Fixed effects | | | | | | | |
| $R^2$ | 0.1100 | | 0.1099 | | 0.1089 | | 0.1058 | |
| N | 1,763,932 | | 1,763,932 | | 1,763,932 | | 1,342,012 | |

*** $p < 0.001$, ** $p < 0.01$, * $p < 0.05$

**Table S34. Pearson correlations between logged perplexity and JIF (natural and social sciences).**

| Model | r | 95% CI | N |
| --- | --- | --- | --- |
| OLMo-1B | 0.0420*** | [0.0405, 0.0435] | 1,774,065 |
| OLMo-7B | 0.0378*** | [0.0364, 0.0393] | 1,774,065 |
| Llama-3-8B | 0.0387*** | [0.0372, 0.0402] | 1,774,065 |
| DeepSeek-LLM-7B | 0.0489*** | [0.0472, 0.0505] | 1,351,531 |

*** $p < 0.001$, ** $p < 0.01$, * $p < 0.05$

**Table S35. Quadratic regression between logged perplexity and citation (natural and social sciences).**

| Variable | OLMo-1B | | OLMo-7B | | Llama-3-8B | | DeepSeek-LLM-7B | |
|---|---|---|---|---|---|---|---|---|
| Log (PPL) | -1.0682 *** (0.0222) | -0.7383 *** (0.0224) | -1.0816 *** (0.0232) | -0.7500 *** (0.0235) | -0.9044 *** (0.0209) | -0.6506 *** (0.0212) | -0.4714 *** (0.0149) | -0.3060 *** (0.0151) |
| Abstract length | | controlled | | controlled | | controlled | | controlled |
| Field | | controlled | | controlled | | controlled | | controlled |
| Publication date | | controlled | | controlled | | controlled | | controlled |
| $R^2$ | 0.002 | 0.024 | 0.002 | 0.024 | 0.001 | 0.023 | 0.001 | 0.020 |
| Adj. $R^2$ | 0.002 | 0.024 | 0.002 | 0.024 | 0.001 | 0.023 | 0.001 | 0.020 |
| N | 1,784,196 | 1,773,985 | 1,784,196 | 1,773,985 | 1,784,196 | 1,773,985 | 1,358,755 | 1,349,163 |

*** $p < 0.001$, ** $p < 0.01$, * $p < 0.05$

**Table S36. Pearson correlations between logged perplexity and citation for high impact journals (natural and social sciences).**

| Model | Top 5% by JIF | | | Top 10% by JIF | | |
|---|---|---|---|---|---|---|
| | r | 95% CI | n | r | 95% CI | n |
| OLMo-1B | -0.1178*** | [-0.1243, -0.1114] | 89,202 | -0.0791*** | [-0.0837, -0.0744] | 177,480 |
| OLMo-7B | -0.1258*** | [-0.1322, -0.1193] | 89,202 | -0.0864*** | [-0.0910, -0.0818] | 177,480 |
| Llama-3-8B | -0.1244*** | [-0.1309, -0.1180] | 89,202 | -0.0832*** | [-0.0878, -0.0786] | 177,480 |
| DeepSeek-LLM-7B | -0.1201*** | [-0.1276, -0.1127] | 67,664 | -0.0809*** | [-0.0862, -0.0756] | 135,201 |

*** $p < 0.001$, ** $p < 0.01$, * $p < 0.05$

**Table S37. Negative binomial regression results between logged perplexity and interdisciplinary-to-intradisciplinary reference ratio (natural and social sciences).** We identified, for each paper, the number of interdisciplinary and intradisciplinary references. We then performed a negative binomial regression to examine how the rate of interdisciplinary to intradisciplinary references changes with paper perplexity. The number of intradisciplinary references was used as an offset in the regression.

| Model | IRR | 95% CI | N |
|---|---|---|---|
| OLMo-1B | 1.3317*** | [1.3271, 1.3363] | 1,770,425 |
| OLMo-7B | 1.4318*** | [1.4267, 1.4370] | 1,770,425 |
| Llama-3-8B | 1.6247*** | [1.6190, 1.6305] | 1,770,425 |
| DeepSeek-LLM-7B | 1.5635*** | [1.5576, 1.5695] | 1,344,979 |

*** $p < 0.001$, ** $p < 0.01$, * $p < 0.05$

**Table S38. Negative binomial regression results between logged perplexity and interdisciplinary-to-intradisciplinary citation ratio (natural and social sciences).** As in Table S37, but for citations rather than references.

| Model | IRR | 95% CI | N |
|---|---|---|---|
| OLMo-1B | 1.1813*** | [1.1760, 1.1867] | 1,387,792 |
| OLMo-7B | 1.2349*** | [1.2291, 1.2407] | 1,387,792 |
| Llama-3-8B | 1.3146*** | [1.3086, 1.3207] | 1,387,792 |
| DeepSeek-LLM-7B | 1.2766*** | [1.2702, 1.2831] | 1,045,043 |

*** $p < 0.001$, ** $p < 0.01$, * $p < 0.05$

Table S39. Logistic regression of top 5% IF on logged perplexity (arts and humanities).

| Variable | OLMo-1B | | OLMo-7B | | Llama-3-8B | | DeepSeek-LLM-7B | |
|---|---|---|---|---|---|---|---|---|
| | OR | 95% CI | OR | 95% CI | OR | 95% CI | OR | 95% CI |
| Log (Perplexity) | 0.23 *** | [0.21, 0.24] | 0.23 *** | [0.21, 0.25] | 0.28 *** | [0.26, 0.30] | 0.32 *** | [0.29, 0.34] |
| N | 74,517 | | 74,517 | | 74,517 | | 57,484 | |
| $R^2$ | 0.044 | | 0.043 | | 0.034 | | 0.028 | |

*** $p < 0.001$, ** $p < 0.01$, * $p < 0.05$

**Table S40. Logistic regression of top 5% IF on logged perplexity in each field (arts and humanities fields), with different models as fixed effect.**

| Field | OR | 95% CI | $R^2$ | N |
|---|---|---|---|---|
| Arts & Humanities, Interdisciplinary | 0.21*** | [0.19, 0.23] | 0.047 | 45,753 |
| History & Archaeology | 0.74*** | [0.67, 0.82] | 0.002 | 57,204 |
| Literature & Language | 0.53*** | [0.49, 0.58] | 0.008 | 68,807 |
| Philosophy & Religion | 0.21*** | [0.19, 0.24] | 0.040 | 52,655 |
| Visual & Performing Arts | 0.82*** | [0.79, 0.85] | 0.001 | 234,834 |

*** $p < 0.001$, ** $p < 0.01$, * $p < 0.05$

**Table S41. Logistic regression of bottom 5% IF on logged perplexity (arts and humanities).**

| Variable | OLMo-1B | | OLMo-7B | | Llama-3-8B | | DeepSeek-LLM-7B | |
|---|---|---|---|---|---|---|---|---|
| | OR | 95% CI | OR | 95% CI | OR | 95% CI | OR | 95% CI |
| Log (PPL) | 5.19 *** | [4.79, 5.62] | 5.14 *** | [4.74, 5.58] | 3.91 *** | [3.61, 4.23] | 3.55 *** | [3.24, 3.88] |
| N | 74,517 | | 74,517 | | 74,517 | | 57,482 | |
| $R^2$ | 0.055 | | 0.054 | | 0.039 | | 0.033 | |

*** $p < 0.001$, ** $p < 0.01$, * $p < 0.05$

**Table S42. Logistic regression of bottom 5% IF on logged perplexity in each field (arts and humanities fields), with different models as fixed effect.**

| Field | OR | 95% CI | $R^2$ | N |
|---|---|---|---|---|
| Arts & Humanities, Interdisciplinary | 2.21*** | [1.99, 2.45] | 0.012 | 45,753 |
| History & Archaeology | 1.92*** | [1.75, 2.11] | 0.008 | 57,204 |
| Literature & Language | 4.96*** | [4.56, 5.40] | 0.052 | 68,807 |
| Philosophy & Religion | 2.86*** | [2.61, 3.14] | 0.022 | 52,655 |
| Visual & Performing Arts | 11.66*** | [11.12, 12.24] | 0.117 | 234,834 |

*** $p < 0.001$, ** $p < 0.01$, * $p < 0.05$

**Table S43.** Logistic regression analysis of perplexity predicting high and low impact factor publications, with abstract length, field, and publication date controlled (arts and humanities).

| Variable | Dependent variable: is in top 5% JIF | | | | | | | |
|---|---|---|---|---|---|---|---|---|
| | Models | | | | | | | |
| | OLMo-1B | | OLMo-7B | | Llama-3-8B | | DeepSeek-LLM-7B | |
| | OR | 95% CI | OR | 95% CI | OR | 95% CI | OR | 95% CI |
| Log (Perplexity) | 0.75 *** | [0.72, 0.78] | 0.73 *** | [0.71, 0.76] | 0.75 *** | [0.72, 0.77] | 0.78 *** | [0.75, 0.81] |
| Abstract length | Controlled | | | | | | | |
| Field | Fixed effects | | | | | | | |
| Pub. date | Fixed effects | | | | | | | |
| $R^2$ | 0.2658 | | 0.2668 | | 0.2661 | | 0.2692 | |
| N | 65,343 | | 65,343 | | 65,343 | | 49,660 | |
| Variable | Dependent variable: is in bottom 5% JIF | | | | | | | |
| | Models | | | | | | | |
| | OLMo-1B | | OLMo-7B | | Llama-3-8B | | DeepSeek-LLM-7B | |
| | OR | 95% CI | OR | 95% CI | OR | 95% CI | OR | 95% CI |
| Log (Perplexity) | 1.72 *** | [1.66, 1.79] | 1.73 *** | [1.66, 1.80] | 1.58 *** | [1.52, 1.64] | 1.47 *** | [1.41, 1.53] |
| Abstract length | Controlled | | | | | | | |
| Field | Fixed effects | | | | | | | |
| Pub. date | Fixed effects | | | | | | | |
| $R^2$ | 0.1185 | | 0.1185 | | 0.1101 | | 0.1111 | |
| N | 65,343 | | 65,343 | | 65,343 | | 49,660 | |

*** $p < 0.001$, ** $p < 0.01$, * $p < 0.05$

**Table S44. Pearson correlations between logged perplexity and JIF (arts and humanities).**

| Model | r | 95% CI | N |
| --- | --- | --- | --- |
| OLMo-1B | -0.2996*** | [-0.3061, -0.2931] | 74,517 |
| OLMo-7B | -0.2915*** | [-0.2981, -0.2849] | 74,517 |
| Llama-3-8B | -0.2432*** | [-0.2500, -0.2364] | 74,517 |
| DeepSeek-LLM-7B | -0.2346*** | [-0.2423, -0.2269] | 57,484 |

*** $p < 0.001$, ** $p < 0.01$, * $p < 0.05$

Table S45. Pearson correlations between logged perplexity and citation (arts and humanities).

| Model | r | 95% CI | N |
|---|---|---|---|
| OLMo-1B | -0.2144*** | [-0.2212, -0.2076] | 74,881 |
| OLMo-7B | -0.2113*** | [-0.2181, -0.2045] | 74,881 |
| Llama-3-8B | -0.1815*** | [-0.1884, -0.1745] | 74,881 |
| DeepSeek-LLM-7B | -0.1770*** | [-0.1849, -0.1691] | 57,727 |

*** $p < 0.001$, ** $p < 0.01$, * $p < 0.05$

**Table S46. Negative binomial regression results between logged perplexity and interdisciplinary-to-intradisciplinary reference ratio (arts and humanities).** As in Table S37, but for arts and humanities rather than natural and social sciences.

| Model | IRR | 95% CI | N |
| --- | --- | --- | --- |
| OLMo-1B | 0.7650*** | [0.7503, 0.7800] | 71,356 |
| OLMo-7B | 0.8379*** | [0.8218, 0.8543] | 71,356 |
| Llama-3-8B | 0.9551*** | [0.9370, 0.9735] | 71,356 |
| DeepSeek-LLM-7B | 0.8368*** | [0.8186, 0.8554] | 55,191 |

*** $p < 0.001$, ** $p < 0.01$, * $p < 0.05$

**Table S47. Negative binomial regression results between logged perplexity and interdisciplinary-to-intradisciplinary citation ratio (arts and humanities).** As in Table S46, but for citations rather than references.

| Model | IRR | 95% CI | N |
| --- | --- | --- | --- |
| OLMo-1B | 0.9415*** | [0.9147, 0.9691] | 45,355 |
| OLMo-7B | 0.9743† | [0.9465, 1.0030] | 45,355 |
| Llama-3-8B | 1.0277 | [0.9989, 1.0574] | 45,355 |
| DeepSeek-LLM-7B | 0.9803 | [0.9484, 1.0134] | 34,831 |

*** $p < 0.001$, ** $p < 0.01$, * $p < 0.05$, † $p < 0.1$